\documentclass[a4paper,11pt]{article}
\pdfoutput=1

\usepackage{jheppub}
\usepackage{amsmath,amssymb,amsfonts}
\usepackage{graphicx}
\usepackage{cleveref}
\usepackage{srcltx}
\usepackage{bm}
\usepackage{mathdots}
\usepackage{color}
\usepackage{dsfont}
\usepackage{slashed}
\usepackage{physics}
\usepackage{subfigure}
\usepackage{tikz}
\usepackage[normalem]{ulem}
\usepackage{orcidlink}
\usepackage{enumitem}

\newcommand{\al}[1]{\begin{align} #1 \end{align} }
\newcommand{\prn}[1]{ \left(  #1 \right) }
\newcommand{\avg}[1]{\langle #1 \rangle}

\title{Linear Sigma Dark Matter}
\preprint{LCTP-22-06}

\author[a]{Dan Kondo,}\emailAdd{dan.kondo@ipmu.jp}
\author[b]{Robert McGehee\,\orcidlink{0000-0002-9265-0494},}\emailAdd{rmcgehee@umich.edu}
\author[a]{Tom Melia\,\orcidlink{0000-0001-8669-6958},}\emailAdd{tom.melia@ipmu.jp}
\author[a,c,d,1]{and Hitoshi Murayama\,\orcidlink{0000-0001-5769-9471}\note{Hamamatsu Professor}}\emailAdd{hitoshi@berkeley.edu}

\affiliation[a]{Kavli Institute for the Physics and Mathematics of the Universe (WPI), University of Tokyo Institutes for Advanced Study, University of Tokyo, Kashiwa 277-8583, Japan}
\affiliation[b]{Leinweber Center for Theoretical Physics, Department of Physics,\\ University of Michigan, Ann Arbor, MI 48109, USA}
\affiliation[c]{Department of Physics, University of California, Berkeley, CA 94720, USA}
\affiliation[d]{Ernest Orlando Lawrence Berkeley National Laboratory, Berkeley, CA 94720, USA}

\abstract{
We present a model of self-interacting dark matter based on QCD-like theories and inspired by the proximity of $a_0(980\pm 20)$ to the $K\bar{K}(990)$ threshold. Dark matter is comprised of dark pions which self-scatter via the $\sigma$ resonance close to the $\pi\pi$ threshold. While the linear sigma model serves as a qualitative guide, a fully unitary description of the scattering in the strongly coupled regime is given by effective range theory. The introduction of a kinetically mixed dark photon allows the dark pion to either freeze-out or -in. We study the viable parameter space which explains the observed relic abundance while evading all current constraints. Searches for dark matter self interactions at different scales, (in)direct detection signals, and (in)visibly-decaying dark photons will test this model in the near future. 
}
\begin{document}
\maketitle
\flushbottom
\setcounter{page}{2}
\newpage

\section{Introduction}
Gravitational inferences of dark matter distributions are increasing in number and reaching an unprecedented level of precision. Even early on, comparisons of data to those from simulations~\cite{Dubinski:1991bm,Navarro:1995iw,Navarro:1996gj,Dave:2000ar} hinted that dark matter might not be as cold and collisionless as assumed in $\Lambda$CDM. Over the past two decades, some of these hints have persisted and been elevated to the level of ``problems.'' The core-vs-cusp problem, for instance, refers to the cored density profiles observed in both dwarf \cite{Moore:1994yx,Flores:1994gz,Walker:2011zu} and low surface brightness galaxies \cite{deBlok:2001hbg,deBlok:2002vgq,Simon:2004sr, 10.1111/j.1365-2966.2004.07836.x,10.1111/j.1365-2966.2009.15004.x} which have less of a cusp than those from simulations (see \emph{e.g.}~\cite{Tulin:2017ara} for details). Other examples include the too-big-to-fail \cite{Boylan-Kolchin:2011qkt} and diversity \cite{Oman:2015xda} problems. 

These observations have motivated models of self-interacting dark matter (SIDM) as possible resolutions.\footnote{Baryonic feedback may reconcile collisionless cold dark matter with the observed mass distributions in galaxies \cite{Bullock:2017xww}; future surveys such as the Prime Focus Spectrograph (PFS) on the Subaru telescope \cite{PFSTeam:2012fqu} may shed light on these different possibilities.} Indeed, given that the dark matter accounts for over $80\%$ of all known matter in the Universe \cite{Planck:2018vyg}, it seems a credible possibility that it belongs to a dark sector as rich as that of the visible. Various realizations of SIDM have used long-range forces \cite{Spergel:1999mh,Feng:2009hw,Koren:2019iuv,Agrawal:2020lea}, self-heating \cite{Kamada:2017gfc,Chu:2018nki,Kamada:2018hte}, and inelastic scatters \cite{McDermott:2017vyk,Vogelsberger:2018bok} to address the small-scale issues. In recent years, models of SIDM have even started attempting to explain a possible velocity dependence in the dark matter self-interaction cross section. Data from dwarf galaxies to galaxy clusters hint that dark matter self interactions may be larger at smaller velocities \cite{Kaplinghat:2015aga}. Resonant self-interacting dark matter is particularly adept at fitting the inferred velocity-dependent cross section~\cite{Chu:2018fzy,Tsai:2020vpi}.

Dark matter self interactions have traditionally been discussed using explicit models of interactions, such as Yukawa ({\it e.g.} \cite{Tulin:2013teo}) or contact interactions ({\it e.g.} \cite{Spergel:1999mh}). While this method serves the purpose of identifying viable models, many models in fact end up producing similar self interactions. This is because the kinematics of dark matter self-scattering in the present-day universe are limited to very low velocities $v \lesssim 10^{-2} c$. Historically, the same problem arose when people tried to understand the anomalously large, low-energy nucleon-nucleon scattering cross sections. Instead of an explicit model, Hans Bethe proposed a parameterization of the low-energy scattering amplitudes called effective range theory (ERT) \cite{Bethe:1949yr} to avoid redundant discussion. Bethe showed that low-energy scattering amplitudes for any two-body potential could be described by just two parameters for $s$-wave interactions: the scattering length, $\mathfrak{a}$, and the effective range, $\mathfrak{r}_e$. ERT was revisited recently in the context of self-interacting dark matter \cite{Chu:2019awd} and parameters consistent with explaining the small-scale discrepancies in various galaxies and clusters of galaxies were identified. 

\begin{figure*}[t!]
\begin{center}
\includegraphics[width=0.75 \columnwidth]{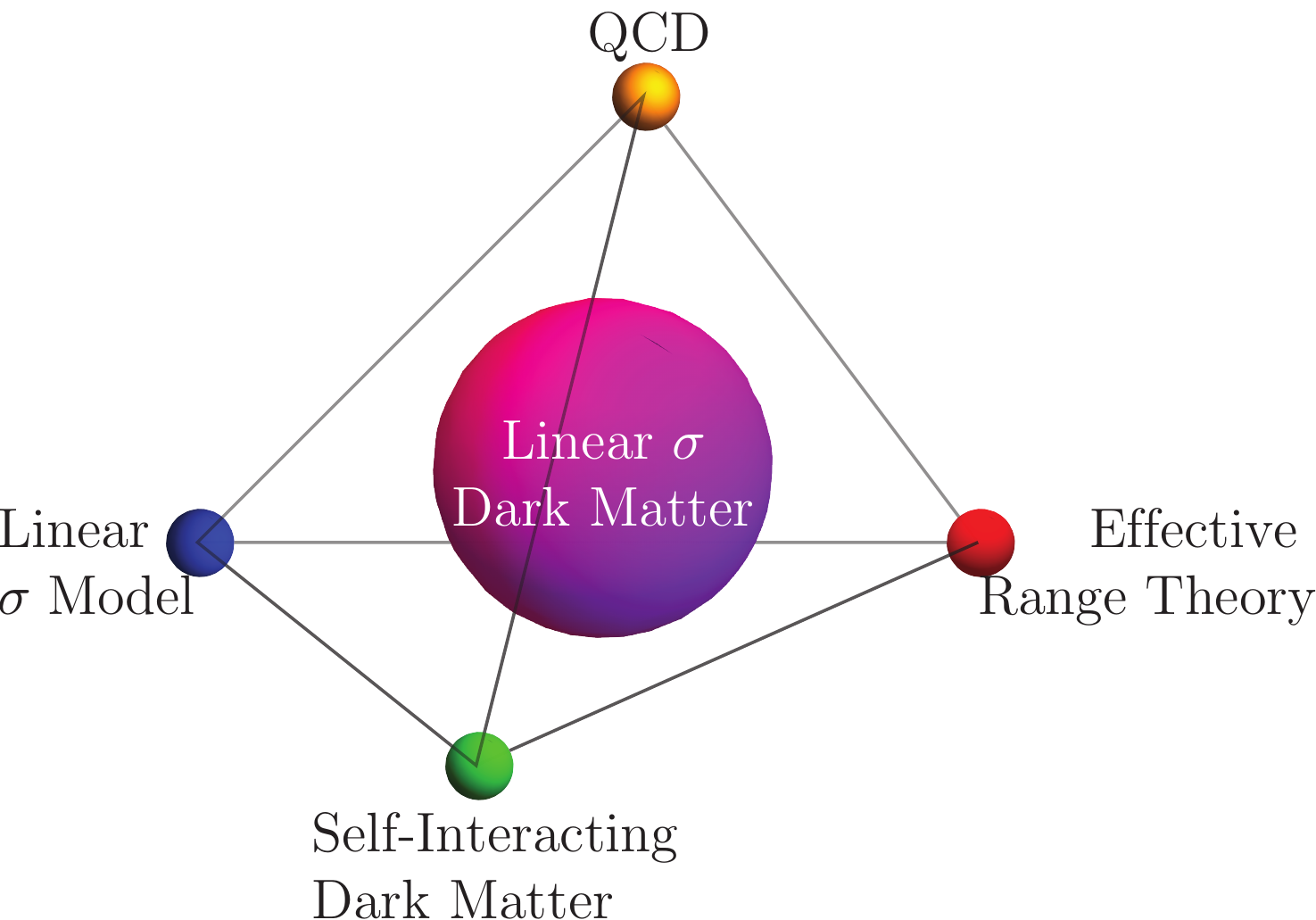}
\caption{\label{fig:schematic} Depiction of the conceptual ingredients which make up linear $\sigma$ dark matter. } 
\end{center}
\end{figure*}

In this paper, we explore a framework for SIDM in which dark matter particles are the pions of a QCD-like theory, focusing on regions of parameter space that can explain the small-scale observations. We find that these regions correspond to resonant dark matter self interactions, providing an explanation for the anomalously large scattering length that an ERT analysis points towards. This is similar to the approach taken in \cite{Tsai:2020vpi}, where a non-linear sigma model (chiral Lagrangian) is used to analyze the pion scattering. However, in the current treatment, we instead use a {\it linear} sigma model (L$\sigma$M) to describe the particle spectrum of the strongly coupled theory. This is inspired by the proximity of $a_0(980\pm 20)$ to the $K\bar{K}(990)$ threshold---a proof of principle that the universe can entertain the sort of resonances we are proposing in the dark sector---and the fact that the L$\sigma$M provides a qualitatively good description of QCD, a point perhaps under-appreciated until recently. A depiction of all these ingredients is given in Fig.~\ref{fig:schematic}.

Via the introduction of a kinetically mixed dark photon, we study the cosmological production of dark matter within such theories, focusing on explicit cases of $SO(N_c)$ and $Sp(2N_c)$ gauge theories. We find viable parameter space in which the dark pions freeze-out or -in to the correct abundance, and we consider possible signals of the models in both indirect and direct detection searches, as well as at  colliders and beam dump experiments. In constructing such realistic models, we note an additional feature of L$\sigma$Ms that is not available in the non-linear models studied previously~\cite{Tsai:2020vpi}, namely the existence of $s$-wave resonances (as opposed to just $p$-wave). This allows for much heavier dark matter which in turn relaxes some phenomenological constraints. 

The remainder of the paper is organised as follows. In Section~\ref{sec:ert}, we review ERT and  perform a scan of ERT parameters that describe dark matter self-interactions to find a best fit to the galaxy data. Section~\ref{sec:lsm0} outlines the framework of the L$\sigma$M and its mapping to the scattering length and effective range parameters of ERT. Section~\ref{sec:models} considers explicit models that reproduce the dark matter relic abundance, as well as various constraints and future experimental sensitivities to the parameter space. We conclude in Section~\ref{sec:conc}. Appendix~\ref{sec:LSM} reviews details of  L$\sigma$Ms that are associated with vector-like gauge theories, and includes a review of the current status of the  L$\sigma$M as applied to QCD. Appendix~\ref{sec:boltz} collects the Boltzmann equations and the formalism that accounts for thermal effects on the dark photon, while Appendix~\ref{sec:xsec} lists cross sections that are used in the calculations throughout the paper.

\section{Effective range theory}
\label{sec:ert}
In this section, we review ERT and perform $\chi^2$ fits to data, leaving the interpretation of the fit parameters within the L$\sigma$M and QCD-like theories to later sections. ERT is an expansion of the phase shift in a power series of the momentum $k$. Since $l$-wave phase shifts are known to be proportional to $k^{2l+1}$ at low momenta, we expand $k^{2l+1} \cot \delta_l$ which is regular as $k\to 0$,
\begin{align}
    k^{2l+1}\cot\delta_{l}(k) = -\frac{1}{\mathfrak{a}_{l}^{2l+1}}+\frac{k^{2}}{2\mathfrak{r}_{e,l}^{2l-1}} + O(k^4).
\end{align}
In fact, Hans Bethe proved that this expansion is always possible for any two-body potential \cite{Bethe:1949yr} for the $s$-wave at low momenta. Solving for the phase shift $\delta_l$, we can identify the scattering amplitude,
\begin{align}
\frac{1}{k} \sin \delta_l e^{i\delta_l}
=\frac{k^{2l}}{-\frac{1}{\mathfrak{a}_{l}^{2l+1}}+\frac{k^{2}}{2\mathfrak{r}_{e,l}^{2l-1}}-ik^{2l+1}}\,,
\end{align}
and the partial-wave cross section
\begin{align}
    \sigma_l(k) &= 4\pi (2l+1) \frac{1}{k^2} \sin^2 \delta_l
    = 4\pi (2l+1) \left|\frac{k^{2l}}{-\frac{1}{\mathfrak{a}_{l}^{2l+1}}+\frac{k^{2}}{2\mathfrak{r}_{e,l}^{2l-1}}-ik^{2l+1}}\right|^2 .
\end{align}
In particular, the $s$-wave amplitude is given by
\begin{align}
\frac{1}{k} \sin \delta_0 e^{i\delta_0}
=\frac{1}{-\frac{1}{\mathfrak{a}}+\frac{\mathfrak{r}_{e}k^{2}}{2}-ik}\ ,
\end{align}
with the partial-wave cross section
\begin{align}
    \sigma_0(k) &= 4\pi \frac{1}{k^2} \sin^2 \delta_0
    = \frac{4\pi \mathfrak{a}^2}{\left|1-\frac{1}{2}\mathfrak{a}\mathfrak{r}_{e}k^{2}+ik\mathfrak{a}\right|^2}\ .
    \label{eq:sigma0}
\end{align}
The thermally averaged cross section for dark matter self scattering (times velocity), $\langle \sigma_0 v \rangle$, can be computed assuming the Maxwell-Boltzmann distribution for relative velocities between dark matter particles in galaxies and clusters of galaxies:
\begin{align}
    P(\vec{v}) = \frac{1}{(\pi v_0^2)^{3/2}} e^{-\vec{v}^2/v_0^2}\ , 
\end{align}
with $\langle v\rangle = 2 v_0/\sqrt{\pi}$. 

\begin{figure}[t]
\begin{center}
\includegraphics[width=0.8\textwidth]{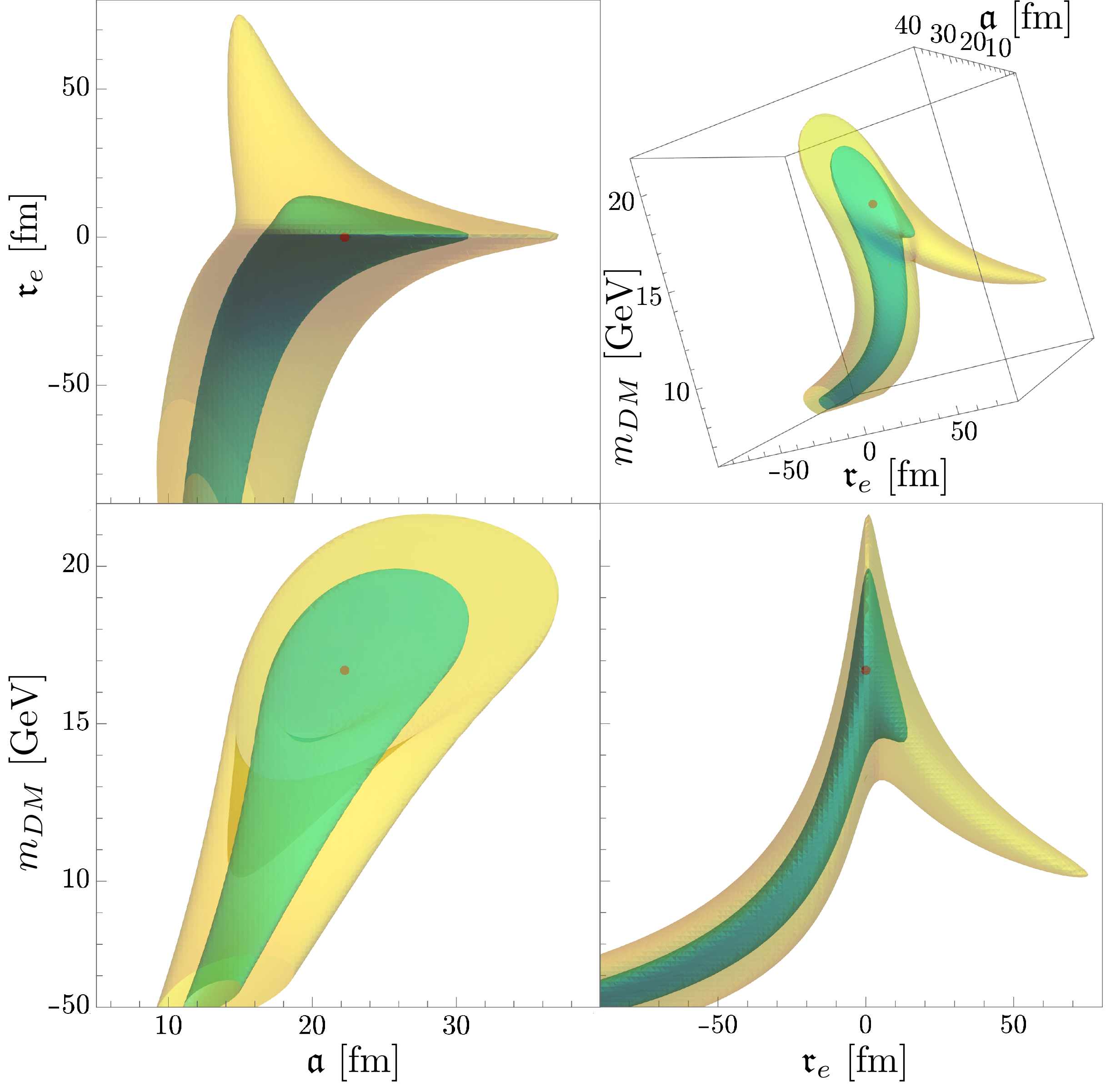}
\caption{\label{fig:scan} $\chi^2$ fit to the data points. The red dot is the best fit point Eq.~\eqref{eq:bestfit}. Green (yellow) regions correspond to 68.27\% (95.45\%) confidence levels for d.o.f.$=3$. The axes are the scattering length $\mathfrak{a} ~\mbox{[fm]}$, effective range $\mathfrak{r}_e~\mbox{[fm]}$, and dark matter mass $m_{\text{DM}}~\mbox{[GeV]}$. The top-right plot shows the 3D regions within the specific confidence levels, while the other three panes show two-dimensional projections. Note that the cross section and fit are insensitive to the simultaneous sign change of $\mathfrak{a}$ and $\mathfrak{r}_e$.}
\end{center}
\end{figure}

\begin{figure}[t]
\begin{center}
\includegraphics[width=0.75\textwidth]{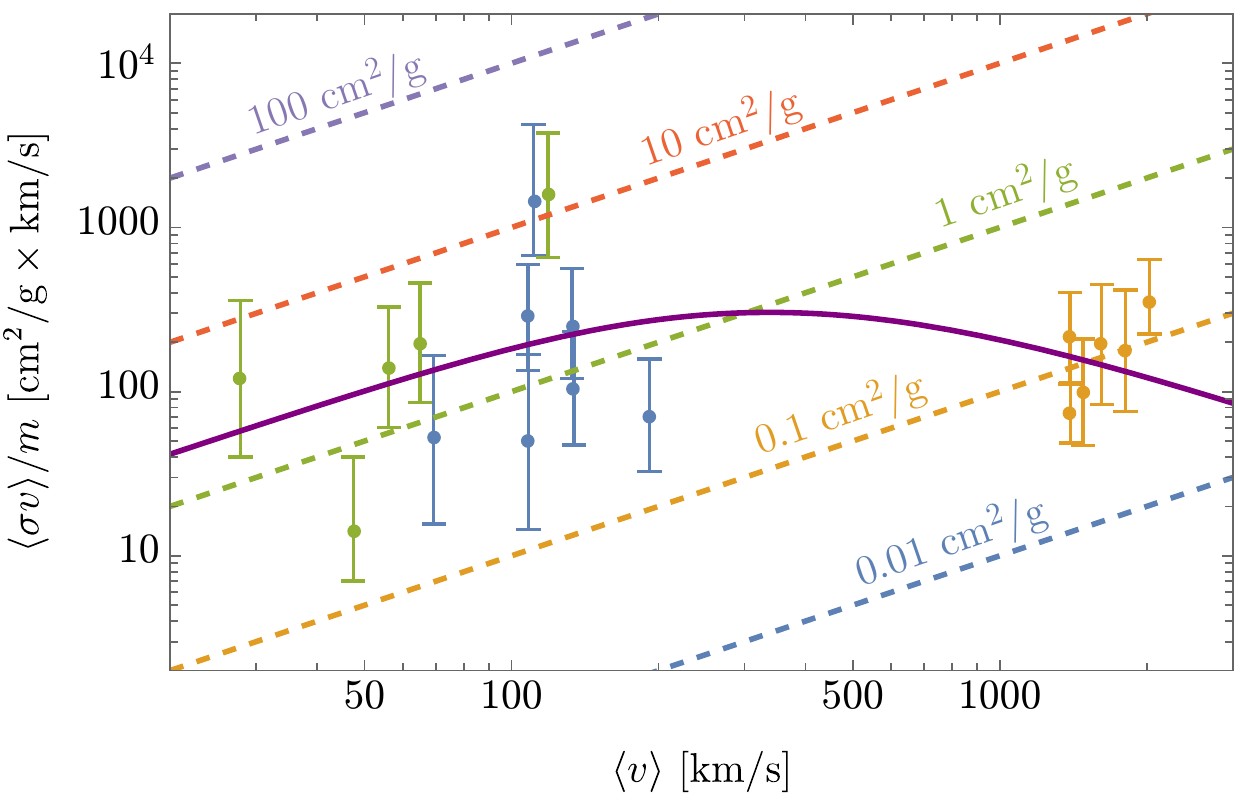}
\caption{\label{fig:bestfit} The best fit curve to the data points from \cite{Kaplinghat:2015aga} with the parameters given in Eq.~\eqref{eq:bestfit}. Here $v$ is the relative velocity, and the dashed lines correspond to constant velocity-independent cross sections.}
\end{center}
\end{figure}

With this ERT cross section, we can determine what values of $\mathfrak{a}$ and $\mathfrak{r}_{e}$ best fit the ``data'' of dark matter self interactions inferred from dwarf galaxies, galaxies, and clusters of galaxies in \cite{Kaplinghat:2015aga}. This extraction is based on semi-analytic approximations which are the subject of further discussion (see, {\it e.g.}\/, \cite {Sagunski:2020spe,Andrade:2020lqq}). We use the data points for illustrative purposes assuming no correlations. We perform a scan over $\left( m_{\text{DM}},\mathfrak{a},\mathfrak{r}_e \right)$, computing the $\chi^2$ fit to the data points. The results are shown in Fig.~\ref{fig:scan}. The best-fit parameters are
\begin{align}
\mathfrak{a}&=22.2~\text{fm}, \nonumber \\
\mathfrak{r}_e&=-2.59 \times 10^{-3}~\text{fm}, \label{eq:bestfit} \\
m_{\text{DM}}&=16.7~\text{GeV} \,, \nonumber
\end{align}
with a value of $\chi^2$ per degree of freedom of 2.01; the resulting fit to the data for these particular values is shown in Fig.~\ref{fig:bestfit}. This result is consistent with the existing literature, {\it e.g.}\/, the point S2 in \cite {Chu:2018fzy} and the point S1 in \cite {Chu:2019awd}. There is a long tail of reasonable fits down to smaller $\mathfrak{r}_e$ in Fig.~\ref{fig:scan}. We confirmed analytically that the tail persists indefinitely with the scaling $m_{\text{DM}} \propto |\mathfrak{r}_e|^{-2/5}$, $\mathfrak{a} \propto |\mathfrak{r}_e|^{-1/5}$. We will comment on the tail later in the context of a L$\sigma$M.

We naively expect the scattering length $\mathfrak{a}$ to be about the Compton wavelength of dark matter-- $\frac{\hbar}{m_{\text{DM}}c}\sim 0.015$~fm for the best-fit mass that was found--while the actual best-fit value is about 20~fm. An interesting question is why it is so large, about 1000 times the Compton wavelength. In fact, a similar phenomenon is observed for the iso-singlet channel for $n p$ scattering: $\mathfrak{a}=-23.712 \pm 0.013$~fm \cite{Noyes:1972xkg}, more than a hundred times larger than the Compton wavelength. This is understood as a consequence of the near-threshold bound state of deuteron, a pole in the scattering amplitude for $k$ along the positive imaginary axis $k=i\kappa$. Similarly, the scattering length for the iso-triplet channel of $nn$ scattering, $\mathfrak{a}\simeq -17$~fm, is also anomalously large. This is understood as a pole in the scattering amplitude for a complex $k$ corresponding to a ``virtual state.'' We discuss a physical interpretation of the large enhancement in the scattering length $\mathfrak{a} \gg \frac{\hbar}{m_{\text{DM}}c}$ for the best-fit parameters in Eq.~\eqref{eq:bestfit} within a L$\sigma$M as an effective description of dark, QCD-like gauge theories in the next section.

\section{Linear sigma model}
\label{sec:lsm0}
We use the L$\sigma$M to study the self interactions of dark matter for two reasons. First, we would like to understand what the ERT parameters from the previous section mean physically. For phenomenological purposes, the ERT parameters are all we need for a perfectly unitary and self-consistent description of the self interaction. However, we would like to gain insight into the underlying dynamics that leads to such parameters. As we will see, the necessary ERT parameters actually correspond to a $\sigma$-like resonance just above threshold or a bound state just below threshold in $\pi\pi$-like scattering. The second reason is because the L$\sigma$M actually serves as a qualitatively correct description of QCD. This point has perhaps been underappreciated in the community. In fact, the $SU(3)$ L$\sigma$M predicts a nonet of $0^+$ states which are all now considered well established experimentally (see Appendix~\ref{sec:LSM}). Therefore, the L$\sigma$M allows us to believe that the necessary ERT parameters actually correspond to a $0^+$ resonance or bound state in $\pi\pi$ scattering in QCD-like theories. In the next section, we demonstrate that $Sp(2N_c)$ or $SO(N_c)$ QCD-like theories indeed lead to phenomenologically attractive models of self-interacting dark matter. 

Admittedly, the L$\sigma$M should not be regarded as a quantitatively accurate description of QCD dynamics since the expected self-coupling, $\lambda$, is large:
\begin{align}
    \lambda \approx \frac{(4\pi)^2}{N_c}\ ,
    \label{eq:NDA}
\end{align}
based on naive dimensional analysis (NDA) and the $1/N_c$ expansion. For modest $N_c$, this invalidates perturbative calculations and the L$\sigma$M description cannot be trusted quantitatively. Yet, the L$\sigma$M does allow us to qualitatively identify the physical meaning of the ERT parameters in terms of the $s$-channel $\sigma$ exchange in $\pi \pi$ scattering, as we show in this section.

We start with the Lagrangian
\begin{align}
\mathcal{L}
&=\frac{1}{2} (\partial_\mu \phi_i)(\partial^\mu \phi_i) - V(\phi)\,,
\qquad
V(\phi) = -b\phi_N-\frac{\mu^2}{2} \phi_i\phi_i + \frac{\lambda}{4} (\phi_i \phi_i)^2,
\end{align}
where $i=1,\dots, N$. This is the $SO(N)/SO(N-1)$ L$\sigma$M plus an explicit symmetry breaking term $-b\phi_N$ which physically corresponds to finite quark mass in QCD-like theories. Writing $\phi_i = (\pi_1,\dots, \pi_{N-1}, v+\sigma)$ and requiring the $\sigma$ tadpole to vanish, we have 
\begin{align}
\mathcal{L}
&=\frac{1}{2} (\partial_\mu \Vec{\pi})\cdot(\partial^\mu \Vec{\pi}) +\frac{1}{2} (\partial_\mu \sigma) (\partial^\mu \sigma) - V(\Vec{\pi}, \sigma)\,,
\\
V(\Vec{\pi}, \sigma) &=
\frac{1}{2}m_{\pi}^{2}\Vec{\pi}^2
+\frac{\lambda}{4}(\Vec{\pi}^2)^{2}
+\frac{1}{2}m_{\sigma}^{2}\sigma^{2}
+\lambda v\sigma^{3}
+\frac{\lambda}{4}\sigma^{4}
+\lambda v\Vec{\pi}^2\sigma
+\frac{\lambda}{2}\Vec{\pi}^2\sigma^{2}
\,,
\end{align}
where 
\begin{align}
    m_\sigma^2 &= 3\lambda v^2 - \mu^2, \qquad
    m_\pi^2 = \lambda v^2 - \mu^2 .
    \label{eq:masses}
\end{align}

Calculating the $\pi_i\pi_j\to\pi_k\pi_l$ scattering amplitude at tree level, we find 
\begin{align}
    \mathcal{M}_\text{tree} \bigl(\pi_i\pi_j\to\pi_k\pi_l\bigr) = &-4\lambda^{2}v^{2}\left(\frac{\delta_{ij}\delta_{kl}}{s-m_{\sigma}^{2}} +\frac{\delta_{ik}\delta_{jl}}{t-m_{\sigma}^{2}}+\frac{\delta_{il}\delta_{jk}}{u-m_{\sigma}^{2}}\right) \nonumber \\ &-2 \lambda\left(\delta_{ij}\delta_{kl}+\delta_{ik}\delta_{jl}+\delta_{il}\delta{jk}\right).
\end{align}
In the non-relativistic limit, we can separate the center-of-mass motion and discuss only the relative motion between two particles. In the center-of-mass frame of two identical particles, the relative momentum $\vec{p}$ is given by
\begin{align}
    \vec{p} &= \frac{1}{2}(\vec{p}_1 - \vec{p}_2)=\vec{p}_1=-\vec{p}_2, \\
    s &= 4m_\pi^2 + 4\vec{p}^2 ,
\end{align}
while $t, u\sim\mathcal{O}(p^2)$. We will be interested in the situation where
\begin{equation}
    m_\sigma = (2+\varepsilon)\,m_\pi\qquad\quad \bigl(|\varepsilon|\ll 1\bigr)\,.
    \label{eq:varepsilon}
\end{equation}
In this case, the tree level amplitude is dominated by $s$-channel $\sigma$ exchange. Moreover, loop corrections are generically suppressed by powers of $\frac{p}{m_\pi}$ in the non-relativistic limit, and are only significant when accompanied by factors of $\frac{1}{\varepsilon}$ from $s$-channel $\sigma$ propagators. These loop corrections are re-summed into a momentum-dependent width:
\begin{equation}
    \mathcal{M} \bigl(\pi_i\pi_j\to\pi_k\pi_l\bigr) \simeq -\delta_{ij}\delta_{kl}\, \frac{4\lambda^{2}v^{2}}{s-m_{\sigma}^{2}+im_\sigma\Gamma(p)} 
    \simeq \delta_{ij}\delta_{kl}\, \frac{\lambda^{2}v^{2}}{\varepsilon m_\pi^2 -p^2 -\frac{i}{4}m_\sigma\Gamma(p)} \,,
\end{equation}
where
\begin{equation}
    m_\sigma\Gamma(p) = (N-1) \,\frac{\lambda^2 v^2}{4\pi}\frac{p}{m_\pi} \,.
\end{equation}

When all $(N-1)$ pions are present in the halo, the total scattering rate via $s$-channel $\sigma$ exchange is determined by the $SO(N-1)$ singlet channel, $(\pi\pi)_\sigma \to (\pi\pi)_\sigma$, with $|(\pi\pi)_\sigma\rangle = \frac{1}{\sqrt{N-1}}\sum_{i=1}^{N-1}|\pi_i\pi_i\rangle$. We have 
\begin{equation}
    \mathcal{M} \bigl((\pi\pi)_\sigma \to (\pi\pi)_\sigma\bigr) 
    \simeq -\frac{4(N-1)\lambda^2 v^2}{4m_\pi^2 + 4p^2 - m_\sigma^2 + i m_\sigma \Gamma(p)}
    \simeq 
    \frac{(N-1)\,\lambda^{2}v^{2}}{\varepsilon m_\pi^2 -p^2 -\frac{i}{4}m_\sigma\Gamma(p)}\ 
\end{equation}
and the $s$-wave elastic cross section is
\begin{align}
    \sigma_0 &= \frac{1}{2s} \frac{1}{8\pi} \left| {\cal M} \right|^2.
\end{align}

Comparing to the ERT cross section Eq.~\eqref{eq:sigma0}, we identify 
\begin{align}
    {\cal M} &= \frac{16\pi m_\pi \mathfrak{a}}{1-\frac{1}{2}\mathfrak{ar}_e k^2 + i k \mathfrak{a}}\ .
    \label{eq:identification}
\end{align}
With the parameter relations from Eq.~\eqref{eq:masses}, we find the ``dictionary,''
\begin{equation}
    \mathfrak{a} = -\frac{(N-1)\lambda^2 v^2}{16\pi m_\pi^3 \varepsilon} 
    \simeq -\frac{3(N-1)\lambda}{32\pi m_\pi \varepsilon} \, ,
    \qquad
    \mathfrak{r}_e = -\frac{32\pi m_\pi}{(N-1)\lambda^2 v^2}
    \simeq -\frac{64\pi }{3(N-1)\lambda m_\pi} \, .
    \label{eq:ERTvsLSM}
\end{equation}
We see that the scattering length $\mathfrak{a}$ is parametrically enhanced by $\frac{1}{\varepsilon}$. As expected, the $\sigma$ particle with mass $m_\sigma = (2+\varepsilon)\,m_\pi$ is a near-threshold resonance ($\mathfrak{a}<0$, $\mathfrak{r}_e<0$) when $\varepsilon>0$, or a bound state ($\mathfrak{a}>0$, $\mathfrak{r}_e<0$) when $\varepsilon<0$. 

It is clear from the expression of $\mathfrak{a}$ in Eq.~\eqref{eq:ERTvsLSM} that two factors can enhance the scattering length: a strong coupling, $\lambda\gg 1$, or a near-threshold mass, $\varepsilon \ll 1$. For example, if we choose the NDA estimate $\lambda=(4\pi)^2/N_c$ Eq.~\eqref{eq:NDA} with $N_c=2$, $m_\pi = 16.7$~GeV, and $N=6$, as expected in the the minimal $Sp(2)$ model with $N_f=2$ (see Appendix~\ref{sec:SpLsM}), we reproduce a near-best fit Eq.~\eqref{eq:bestfit} with $\mathfrak{r}_e = -0.0020$~fm and $\varepsilon = 0.0063$. The small $\varepsilon$ seems incidental ({\it i.e.} requiring fine tuning), but this percent-level coincidence does occur in QCD (see Appendix~\ref{sec:SU3LsM}).

There is an $SU(3)$ nonet of light $0^+$ resonances \cite{ParticleDataGroup:2020ssz} that can be identified with the scalar bosons in the $SU(3)$ L$\sigma$M (see Appendix~\ref{sec:SU3LsM}). They can be interpreted as $qq\bar{q}\bar{q}$ states or molecules of pseudoscalar mesons. Among them, $f_{0}(980)$ and $a_{0}(980)$ are basically kaon molecules. Within the experimental uncertainties, $m(f_0) \approx m(a_0) \approx 2m_K$ with percent-level accuracy; see Fig.~\ref{fig:sigmas}. A lattice QCD simulation also shows such a near-threshold behavior is possible \cite{Molina:2018otc} (see \cite{Briceno:2016mjc,Briceno:2017qmb} also).

The long tail to larger $|\mathfrak{r}_e|$ in the $\chi^2$ fit corresponds to fixed $\varepsilon$ and weaker coupling such that $m_\pi \sim \lambda^{2/3}$ and $\mathfrak{a} \sim \lambda^{1/3}$. This may be achieved for large $N_c$. For Strongly Interacting Massive Particle (SIMP) dark matter with dark pions freezing-out via $3 \rightarrow 2$ annihilations due to the Wess--Zumino term in the chiral Lagrangian \cite{Hochberg:2014kqa}, this tail provides the desired velocity dependence even for the $m_\pi \sim 300$~MeV required for the scenario.

\section{Explicit models}
\label{sec:models}
Having established that the L$\sigma$M fits the dark matter self interaction data well, we now detail two explicit L$\sigma$Ms which successfully reproduce the dark matter relic abundance. In the first, we gauge a dark $Sp(2N_c)$ and include $N_f=2$ fermions in the fundamental representation in the dark sector. In the second, we instead gauge a dark $SO(N_c)$ and include $N_f=2$ fermions in the vector representation. In both models, we additionally introduce a $U(1) '$ and its corresponding dark photon, $\gamma'$, which kinetically mixes with the Standard Model (SM) photon via
\begin{align}
    \mathcal{L} \supset \frac{\epsilon}{2} F^{\mu\nu}F'_{\mu\nu} \, ,
\end{align}
where $F_{\mu\nu}$ and $F'_{\mu\nu}$ are the SM and dark photon field strengths. This vector portal allows dark matter to freeze-out (or -in) and effectively decouples the origin of the relic abundance from the origin of the sizeable dark matter self interactions.\footnote{Please note the notation for the kinetic mixing $\epsilon$ as opposed to the degree of degeneracy $\varepsilon$ in Eq.~\eqref{eq:varepsilon}.} As we shall see, we are able to achieve the relic abundance via either freeze-out or freeze-in mechanisms in the $Sp(2N_c)$ model. The more minimal $SO(N_c)$ model, however, only permits the latter mechanism.

\subsection{\texorpdfstring{$Sp(2N_c)$}{Sp(2Nc)}}
The first model we consider is a gauged $Sp(2N_c)$ with $N_f=2$ dark quarks ({\it i.e.}\/, four Weyl fermions) in the fundamental representation. The size of the gauge group $N_c$ determines the strong coupling $\lambda$ in the L$\sigma$M; see Eq.~\eqref{eq:NDA} for the NDA estimate. In the massless quark limit, it has an $SU(4) \simeq SO(6)$ flavor symmetry, which spontaneously breaks to $Sp(4) \simeq SO(5)$ by the quark bi-linear condensate $\langle q_i q_j \rangle \propto J_{ij}$ $(i,j=1, \cdots, 4)$. The low-energy physics is thus described by the $SO(6)/SO(5)$ L$\sigma$M (see Appendix~\ref{sec:SpLsM}). We introduce a degenerate mass for both flavors, reducing the original symmetry to $Sp(4)$. Additionally, we charge the four dark quarks under $U(1)'$ as $+\frac{1}{2},+\frac{1}{2},-\frac{1}{2},-\frac{1}{2}$, leaving an exact $U(2)$ symmetry.\footnote{There is a conjecture that a theory of quantum gravity would not allow for an exact global symmetry, see {\it e.g.}\/,  \cite{MISNER1957525,Polchinski:2003bq,Banks:2010zn,Harlow:2018tng,Harlow:2018jwu}. The exact $U(2)$ symmetry here can be justified by gauging $U(2)$ and breaking $SU(2)$ by a doublet Higgs, leaving an exact custodial $SU(2)$ together with the unbroken gauged $U(1)'$.}

Among the $_{4}C_{2}=6$ quark pairs, we identify one as $\sigma$ and the remaining five as $\pi$s (here and throughout, $\pi$ will refer to \emph{dark} pions). Two of the five $\pi$s have charge $+1$ and $-1$ and the remaining three are neutral as an isotriplet of $SU(2)$. Since they are the lightest states with non-trivial quantum numbers under the exact $SU(2)$, they are stable. Depending on the mass splitting between the charged and neutral pions, dark matter may therefore be comprised of all five states or just the lightest neutral three. This mass splitting, as in the SM, is due to QED corrections and is of the order $\Delta m_\pi^2 \simeq 4\pi \alpha' f_\pi^2 \approx 22.2 \alpha'  \text{ GeV}^2$, since $f_\pi\approx\frac{m_\pi}{4\pi}$. 

Whether or not direct detection can detect this dark matter depends on whether or not (a subset of) it is charged under $U(1)'$. To understand the final relative abundances of the charged and neutral dark pions, 
we first estimate the temperature at which the pion-changing process $\pi^+\pi^{-}\rightarrow \pi^0_i\pi^0_i$ decouples, $T_d$. The cross section for this process is roughly:
\begin{align}
    \sigma(\pi^+\pi^{-}\rightarrow \pi^0_i\pi^0_i)\simeq \frac{1}{32\pi f_\pi^2}.
\end{align}
This process decouples when its rate becomes comparable to Hubble, i.e. $\avg{\sigma v}n_{\pi^+} \sim H$. Since this happens after the relic abundance is set and the dark matter is non-relativistic, $v \sim \sqrt{T/m_\pi}$ and $n_{\pi^+} = 1/5 Y_\text{DM} s$. Solving for $T_d$, we find
\al{
T_d \approx \prn{120 \sqrt{10} \frac{\sqrt{g_\ast}}{g_{\ast s}Y_\text{DM}} \frac{f_\pi^2 \sqrt{m_\pi}}{M_{Pl}}}^{2/3} = 6.2 \times 10^{-4} \text{ GeV}.
}
In the second equality, we have set the dark matter mass to the best fit value from Eq.~\eqref{eq:bestfit}.With the decoupling temperature in hand, it is simple enough to estimate the final relative abundances of the different pion species. If $T_d \lesssim \Delta m_\pi^2/2m_\pi$, then the forward process $\pi^+\pi^{-}\to \pi^0_i\pi^0_i$ dominates over the backward before decoupling and we can expect all pions to be neutral. Since we have set the dark matter mass to the best-fit value, we find that dark matter is comprised entirely of neutral pions as long as
\al{
\alpha' \gtrsim 9.3 \times 10^{-4} 
\label{eq:alphapcond}
}
As we will see in the following sections, dark matter in this model is comprised of all neutral pions when freeze-out sets the relic abundance, but comprised of both charged and neutral pions when its relic abundance is instead set by freeze-in.

\subsubsection{Relic abundance from freeze-out}
\begin{figure*}[t!]
\begin{center}
\includegraphics[width=0.75 \columnwidth]{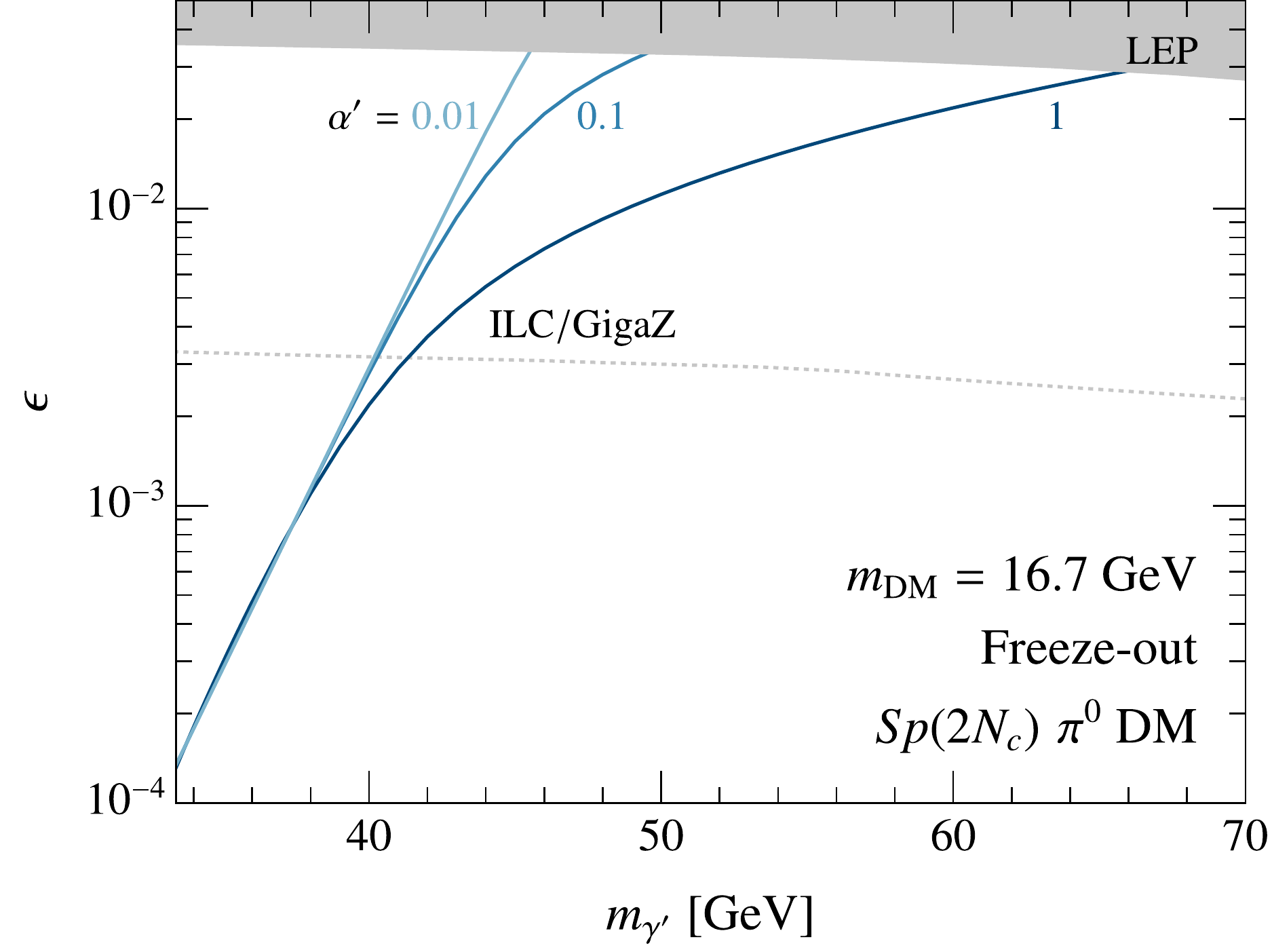}
\caption{\label{fig:Sp2NFO} Contours of $\alpha '$ on the invisibly-decaying dark photon plane $(m_{\gamma '},\epsilon)$ which predict the observed relic abundance via freeze-out in the $Sp(2N_c)$ model with neutral pion dark matter. Also shown in gray is the bound from LEP \cite{Ilten:2018crw} as well as the projected sensitivity from precision electroweak limits at ILC/GigaZ  \cite{Curtin:2014cca} (dashed).}
\end{center}
\end{figure*}

In the freeze-out scenario, the early-Universe abundances of the dark $\pi$s evolve as follows. At temperatures above their mass, all five $\pi$s are in thermal equilibrium with the SM bath thanks to the vector portal and a not-too-small kinetic mixing. When the bath temperature drops below $m_\pi=16.7 \text{ GeV}$, the slightly heavier charged pions start annihilating to pairs of charged SM particles, $\pi^{+} \pi^{-} \to f \bar{f}$. Once the temperature drops below roughly $T \lesssim m_\pi/20$, these annihilations slow sufficiently for the $\pi$s to freeze-out from the SM bath. However, the strong self interactions among the $\pi$s allow them to continue annihilating amongst each other. This causes a down-scattering of charged dark matter states to neutral ones, so that the remaining relic abundance is entirely comprised of the lighter, stable neutral $\pi$s. This allows the dark matter to evade both direct and indirect detection bounds.

The only viable freeze-out parameter space occurs for $m_{\gamma '} > 2m_\pi$.\footnote{There is viable parameter space slightly below this threshold, but the experimental bounds dramatically change since the dark photon becomes visibly decaying. Thus, there is only a sliver of $m_{\gamma '} < 2m_\pi$ before the required $\epsilon$ is ruled out by LHCb $A' \to \mu^+ \mu^-$ visible searches \cite{Aaij:2019bvg,GrillidiCortona:2022kbq}.} Thus, the freeze-out process is $\pi^{+} \pi^{-} \to \bar{f} f$, where $f$ is a SM charged fermion. These annihilations are $p$-wave and the resulting relic abundance is \cite{Gondolo:1990dk} 
\begin{align}
    \label{eq:FOabnd}
    \Omega_{\text{DM}} h^{2}
    &=\frac{8.53\times 10^{-11}}{\text{GeV}^2}\frac{2x_{f}^{2}}{{g_*}^{1/2}}\frac{1}{\avg{\sigma v}_\text{eff}}  \,,
\end{align}
where $h$ is the Hubble constant in units of 100 km sec$^{-1}$ Mpc$^{-1}$ and $\avg{\sigma v}_\text{eff}$ is the thermally averaged cross section for charged pion annihilations summed over all SM final states.\footnote{Since the charged dark pions are only 2 out of the 5 dark pions, the probability that 2 dark pions encountering one another in a $\pi^+$ and $\pi^-$ pair is only $2/25$. Thus, the effective cross section $\avg{\sigma v}_\text{eff}$ that enters Eq.~\eqref{eq:FOabnd} is smaller than the actual cross section by this same factor, $\avg{\sigma v}_\text{eff}=\prn{2/25} \avg{\sigma v}$.} For numerical evaluations, we use $x_f = 20$, where $x_f=m_{\text{DM}}/T_f$ and $T_f$ is the freeze-out temperature, and $g_* = 75$.

Fig.~\ref{fig:Sp2NFO} shows the resulting values of $(m_{\gamma '},\epsilon)$ for different fixed $\alpha '$ which correctly reproduce the observed dark matter relic abundance. Since we only consider $\alpha' \ge 0.01$, the mass splitting between the charged and neutral pions is large enough to guarantee that dark matter is only comprised of the neutral ones. Since the dark photons in this part of parameter space are invisibly decaying, the most stringent bound on the kinetic mixing comes from LEP \cite{Ilten:2018crw}, as shown in gray. Also shown with a dashed gray line is the projected sensitivity of ILC/GigaZ \cite{Curtin:2014cca} using future precision electroweak limits and it is exciting that the ILC will be able to probe most of the freeze-out regime.

\subsubsection{Relic abundance from freeze-in}
Since freeze-out is only possible over a narrow range of $m_{\gamma '}$, we also consider using the next-simplest mechanism to set the relic abundance: freeze-in \cite{Hall:2009bx,Chu:2013jja}. Freeze-in scenarios are characterized by a dark sector that is initially unpopulated and remains out of equilibrium with the SM bath at all times. In such a situation, SM bath particles slowly and steadily produce dark matter (and perhaps other dark-sector particles) through feeble couplings. 

Which freeze-in processes matter most depends on $m_{\gamma '}/m_\pi$, $\epsilon$, and $\alpha'$. When $m_{\gamma '} \ge 2 m_\pi$ and $\gamma '$ may decay to pairs of charged dark pions, the dominant freeze-in production process is inverse decays of SM fermion pairs into dark photons, followed by their subsequent decay into dark pions.\footnote{Contributions from SM Z decays to pairs of charged dark pions is non-negligible if $m_{\gamma '} \gtrsim 54 \text{ GeV}$; for simplicity, we focus on the parameter space for lighter dark photons.} For lighter $\gamma '$, direct production $\bar{f} f \to \pi^{+} \pi^{-}$ (direct freeze-in), or $\bar{f} f \to \gamma \gamma '$ and $f \gamma \rightarrow f \gamma'$ followed by $\gamma ' \gamma ' \to \pi^{+} \pi^{-}$ (sequential freeze-in) are the dominant processes. For smaller values of $\alpha '$, the first process is the most important while for larger values, the latter two dictate the relic abundance \cite{Hambye:2019dwd}. Regardless, all values of $\alpha '$ relevant for freeze-in are less than the estimate in \eqref{eq:alphapcond}. Thus, both charged and neutral pions make up dark matter and $\frac{2}{5}$ of dark matter may be directly detected as we discuss below.

We turn first to the freeze-in production when $m_{\gamma '} \ge 2 m_\pi$. The calculation is brief and well known and the resulting yield of the dark photons is \cite{Koren:2019iuv}
\al{
\label{eq:heavygamFIyield}
Y_{\gamma'} \approx \frac{3 m_{\gamma'}^2}{2 \pi^2} \frac{\prn{45}^{3/2} M_\text{Pl}}{\sqrt{2}\pi^3} \sum_f \Gamma_{\gamma' \to \bar{f} f} \int_0^\infty \frac{K_1(m_{\gamma'}/T)}{\sqrt{g_\ast}g_{\ast s}T^5},
}
where 
\al{
\Gamma_{\gamma' \to \bar{f} f} = \alpha \epsilon^2 \frac{m_{\gamma'}^2+2m_f^2}{3m_{\gamma'}} \sqrt{1-\frac{4 m_f^2}{m_{\gamma'}^2}}
}
is the partial decay rate of $\gamma '$ to a SM fermion pair. In Eq.~\eqref{eq:heavygamFIyield}, $Y_{\gamma'} = n_{\gamma'}/s_{\text{rad}} $ where $s_\text{rad}=\frac{2 \pi^2}{45}g_{\ast s}T^3$ is the entropy density of the SM bath; $M_\text{Pl}$ is the reduced Planck mass; the sum over $f$ includes all charged SM fermions lighter than the top quark; and $g_\ast$ is the relativistic degrees of freedom in the bath. To arrive at this estimate, it is assumed that $\partial_T g_{\ast s}/g_{\ast s} \ll 1/T$ over the interval of integration, which is a reasonable approximation. Additionally, the integral bounds are well approximated by the interval $m_{\gamma'}/20$ to $20m_{\gamma'}$. 

\begin{figure*}[t!]
\begin{center}
\includegraphics[width=0.75 \columnwidth]{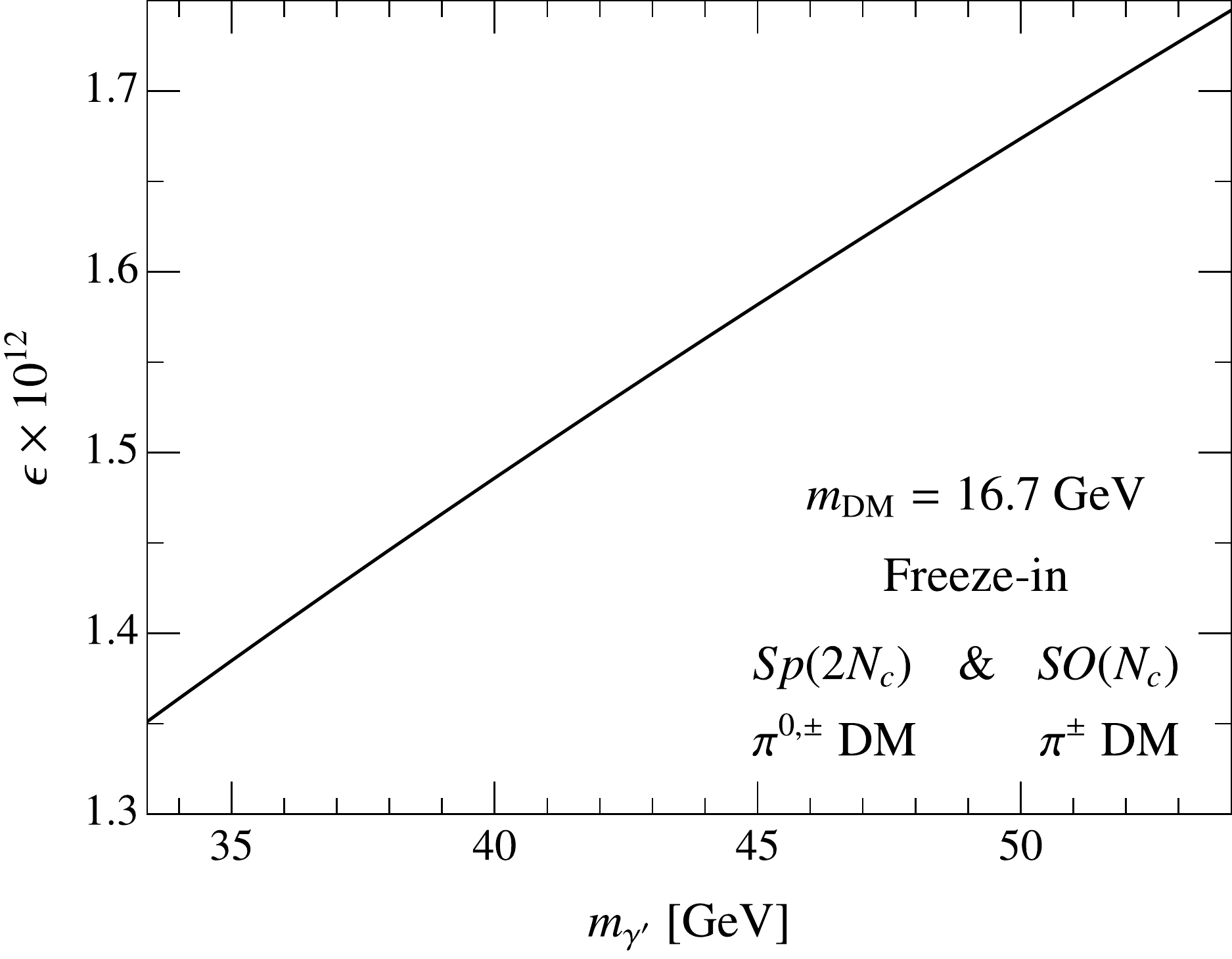}
\caption{\label{fig:heavygamFI} The $\epsilon$ required to achieve the observed relic abundance via freeze-in in both models, for $m_{\gamma '} > 2 m_\pi$. While this is an $\alpha '$-independent result, the assumed range of $\alpha '$ are similar to the values taken in Figs.~\ref{fig:Sp2NlightgamFI} and \ref{fig:SONlightgamFI} to avoid indirect detection bounds. The dark matter mass is assumed to be the best fit value of 16.7~GeV.}
\end{center}
\end{figure*}

After the dark photons freeze-in, they decay to pairs of dark charged pions which then quickly annihilate to pairs of dark neutral pions. Since two dark pions are produced for every dark photon, the value of $\epsilon$ which results in the observed dark matter relic abundance is 
\al{
\epsilon = \sqrt{\frac{\Omega_{\text{DM}} h^2 \rho_\text{crit}/h^2}{2m_\pi \tilde{Y}_{\gamma'} s_0}},
}
where $\Omega_{\text{DM}} h^2 = 0.120$, $\rho_\text{crit}/h^2 = 1.05 \times 10^{-5} \text{ GeV}/\text{cm}^3$, $\tilde{Y}_{\gamma '}=Y_{\gamma '}/\epsilon^2$ (with $Y_{\gamma '}$ given in Eq.~\eqref{eq:heavygamFIyield}), and $s_0=2890 / \text{cm}^3$ \cite{ParticleDataGroup:2020ssz}. The resulting feeble kinetic mixing is shown in Fig.~\ref{fig:heavygamFI}. There are no existing constraints anywhere near the tiny requisite mixings, and no immediate hopes to probe them.

On the other hand, the phenomenology is richer when $m_{\gamma'} < 2 m_\pi$. In this case, the dark photon decays into visible states $\gamma' \rightarrow f \bar{f}$ where $f$ is a SM fermion, and hence is subject to a variety of limits from accelerator experiments and astrophysics. In addition, for not-too-small values of $\alpha' \gtrsim 10^{-13}$, production of dark matter may proceed primarily through the freeze-in of dark photons followed by their subsequent annihilation to dark pions. At the same time, dark photon production from the SM thermal bath may have a resonant effect when the plasma mass of the photon crosses the mass of the dark photon, $\omega_p(T_{\text{res}}) = m_{\gamma'}$ (see Appendix~\ref{sec:thermal} for details). Thus, unlike the simple freeze-in story for $m_{\gamma '} \ge 2 m_\pi$ discussed above, the freeze-in in this parameter space depends on $\alpha'$ in subtle ways which we detail fully in Appendix~\ref{sec:boltz}.

In addition, all such values of $\alpha'$ are below the condition in Eq.~\eqref{eq:alphapcond}. Thus, $2/5$ of dark matter is charged and $3/5$ is neutral and the possibility of directly detecting dark matter opens up. The cross section for charged pion dark matter to scatter off a xenon nucleus at non-relativistic speeds is
\al{
\label{eq:DDsigma}
\sigma_{\pi^\pm \text{Xe}} \approx 16 \pi Z^2 \epsilon^2 \alpha \alpha' \frac{\mu^2}{m_{\gamma '}^4}, 
}
where $Z=54$ is the atomic number of xenon and $\mu$ is the reduced mass of a xenon nucleus and dark pion. The most stringent limit on $16.7 \text{ GeV}$ dark matter, $\chi$, scattering directly off nucleons is $\sigma_{\chi n} \lesssim 8.6 \times 10^{-47} \text{ cm}^2$ \cite{PandaX-4T:2021bab}. This upper limit on the spin-independent cross section is really an upper limit on the dark matter-xenon nucleus cross section, which is related via
\al{
\label{eq:DDbnd}
\sigma_{\chi\text{Xe}} = A^2 \frac{\mu^2}{\mu_{\chi n}^2} \sigma_{\chi n} \lesssim A^2 \frac{\mu^2}{\mu_{\chi n}^2} 8.6 \times 10^{-47} \text{ cm}^2,
}
where $A$ is the atomic mass number of the target xenon isotope, $\mu_{\chi n}$ is the reduced mass of dark matter and a single nucleon, and we have used the current PandaX-4T constraint in the final inequality. We find an upper bound on $\epsilon$ as a function of $m_{\gamma'}$ (for different fixed $\alpha'$) by saturating the bound in Eq.~\eqref{eq:DDbnd} using the cross section for dark pion scattering in Eq.~\eqref{eq:DDsigma} and accounting for only $2/5$ of dark matter being charged. The resulting upper bounds on $\epsilon$ are shown as dashed contours for various $\alpha'$ in Fig.~\ref{fig:SONlightgamFI} and intersect the requisite $\epsilon$ contours which explain the relic abundance at $m_{\gamma '} \approx 50 \text{ MeV}$.

\begin{figure*}[t!]
\begin{center}
\includegraphics[width =0.75 \columnwidth]{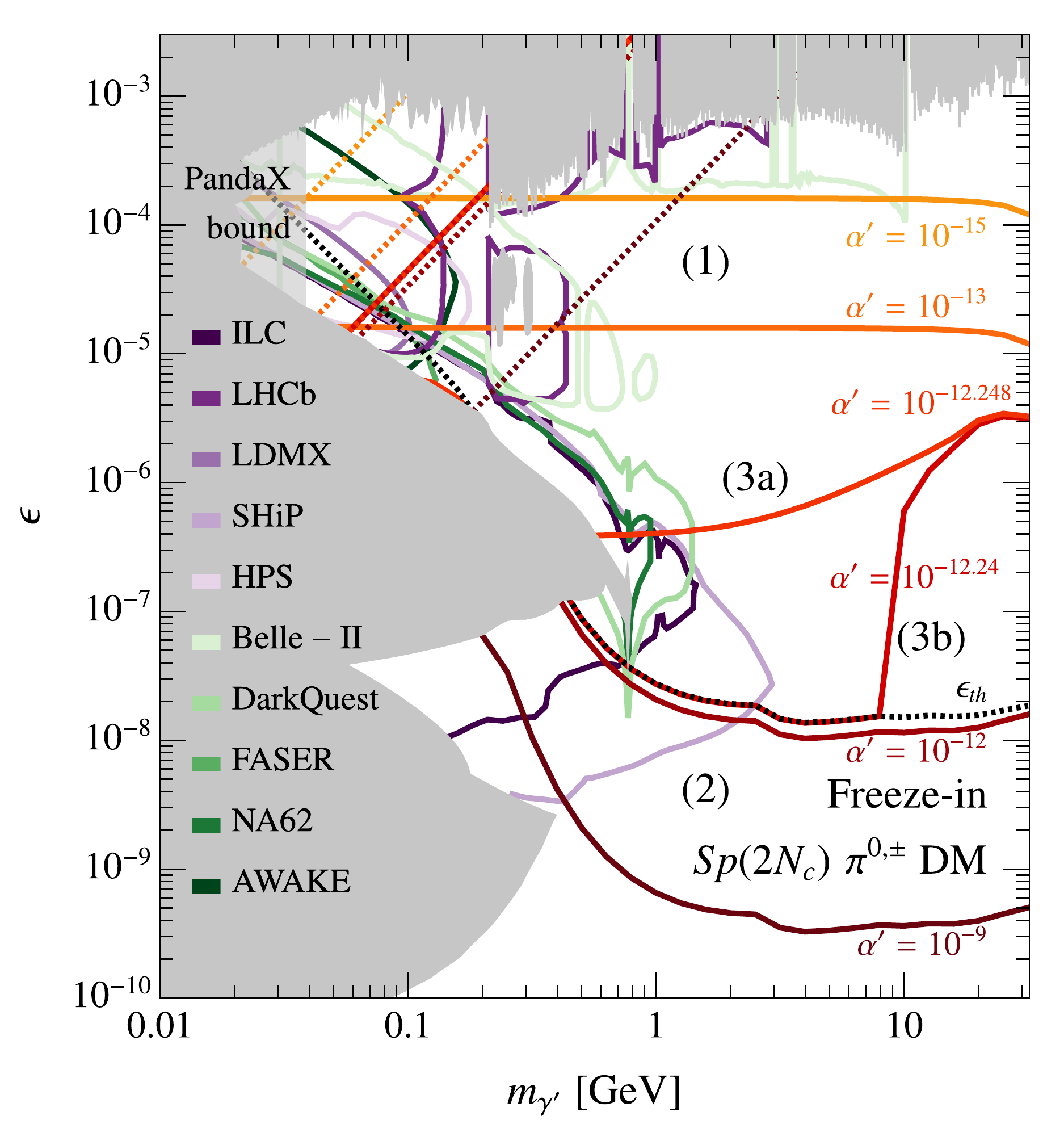}
\caption{\label{fig:Sp2NlightgamFI} Contours of $\alpha '$ on the visibly-decaying dark photon plane $(m_{\gamma '},\epsilon)$ which predict the observed relic abundance via freeze-in in the $Sp(2N_c)$ model with $2/5$ charged and $3/5$ neutral pion dark matter. Constraints are gray \cite{Alexander:2016aln,Chang:2016ntp,Hardy:2016kme,Pospelov:2017kep,Banerjee:2018vgk,Aaij:2017rft,Aaij:2019bvg,Parker:2018vye,Tsai:2019mtm} and projected sensitivities are a variety of purples and greens \cite{Celentano:2014wya,Ilten:2015hya,Alekhin:2015byh,Ilten:2016tkc,Alexander:2016aln,Caldwell:2018atq,Kou:2018nap,Berlin:2018pwi,Berlin:2018bsc,Ariga:2018uku,NA62:2312430,Tsai:2019mtm,Asai:2021ehn,Ferber:2022ewf}. In addition, the latest direct detection bound from PandaX-4T \cite{PandaX-4T:2021bab} is shown as dashed contours with matching colors for each $\alpha'$. (Note the direct detection contours for $\alpha'=10^{-12.248}$ and  $10^{-12.24}$ are overlapping.) The latter results in the light-gray shaded region ruled out by PandaX-4T for $m_{\gamma '} \lesssim 40 \text{ MeV}$. Regions labeled (1), (2), (3a) and (3b) are defined in Appendix~\ref{sec:boltz}. The dark matter mass is assumed to be the best fit value of 16.7~GeV.} 
\end{center}
\end{figure*}

Fig.~\ref{fig:Sp2NlightgamFI} shows contours of $\alpha'$ which result in the observed dark matter relic abundance for $m_{\gamma'} < 2 m_\pi$. Region (1) corresponds to $\alpha'$ contours for which $f \bar{f} \to \pi^+ \pi^-$ is the dominant freeze-in process, while Region (2) corresponds to $\alpha'$ values for which the sequential freeze-in of dark photons followed by their annihilation to dark pions matters most. Region (3) transitions between the two (see Appendix~\ref{sec:boltz} for details). Also shown are the current constraints on visibly-decaying dark photons in gray \cite{Alexander:2016aln,Chang:2016ntp,Hardy:2016kme,Pospelov:2017kep,Banerjee:2018vgk,Aaij:2017rft,Aaij:2019bvg,Parker:2018vye,Tsai:2019mtm} and a variety of projected sensitivities in purples and greens \cite{Celentano:2014wya,Ilten:2015hya,Alekhin:2015byh,Ilten:2016tkc,Alexander:2016aln,Caldwell:2018atq,Kou:2018nap,Berlin:2018pwi,Berlin:2018bsc,Ariga:2018uku,NA62:2312430,Tsai:2019mtm,Asai:2021ehn,Ferber:2022ewf}. In addition to the projections shown, there have been studies of the sensitivity of MUonE \cite{Galon:2022xcl} and  future TeV-scale muon beam dumps to visibly-decaying dark photons \cite{Cesarotti:2022ttv}. Thus, despite dark matter being mostly neutral, this part of parameter space is testable at near-future experiments probing the visible decays of dark photons.

\subsection{\texorpdfstring{$SO(N_c)$}{SO(Nc)} model}
The second model we consider is a gauged $SO(N_c)$ with $N_f=2$ dark quarks ({\it i.e.}\/, two Weyl fermions) in the vector representation. In the massless limit, it has an $SU(2) \simeq SO(3)$ flavor symmetry, which is broken to $SO(2)$ by the quark bi-linear condensate $\langle q_i q_j \rangle \propto \delta_{ij}$ $(i,j=1,2)$. The low-energy physics is described by the $SO(3)/SO(2)$ L$\sigma$M (see Appendix~\ref{sec:SOLsM}). We charge the two dark quarks under a $U(1) '$ as $+\frac{1}{2},-\frac{1}{2}$,  leaving an exact $SO(2) \simeq U(1)'$ symmetry. Among $_{2+1}C_{2}=3$ pairs, one is identified as $\sigma$ and the other two as $\pi^{\pm}$. Since the $\pi^{\pm}$ are the lightest states with non-trivial quantum numbers under the exact $U(1)'$, they are stable. 

The main difference between this model and the previous one is that all of the dark matter is charged instead of neutral. While the previous model was able to accommodate freeze-out for a range of $m_{\gamma '}$, this model cannot since the charged dark matter would scatter too much in direct detection experiments for the necessary values of $\epsilon \sim 10^{-3}$ \cite{XENON:2018voc,PandaX-4T:2021bab}. Thus, we focus only on the freeze-in mechanism for this model to demonstrate a viable parameter space in which the relic abundance is explained.

\begin{figure*}[t!]
\begin{center}
\includegraphics[width=0.75 \columnwidth]{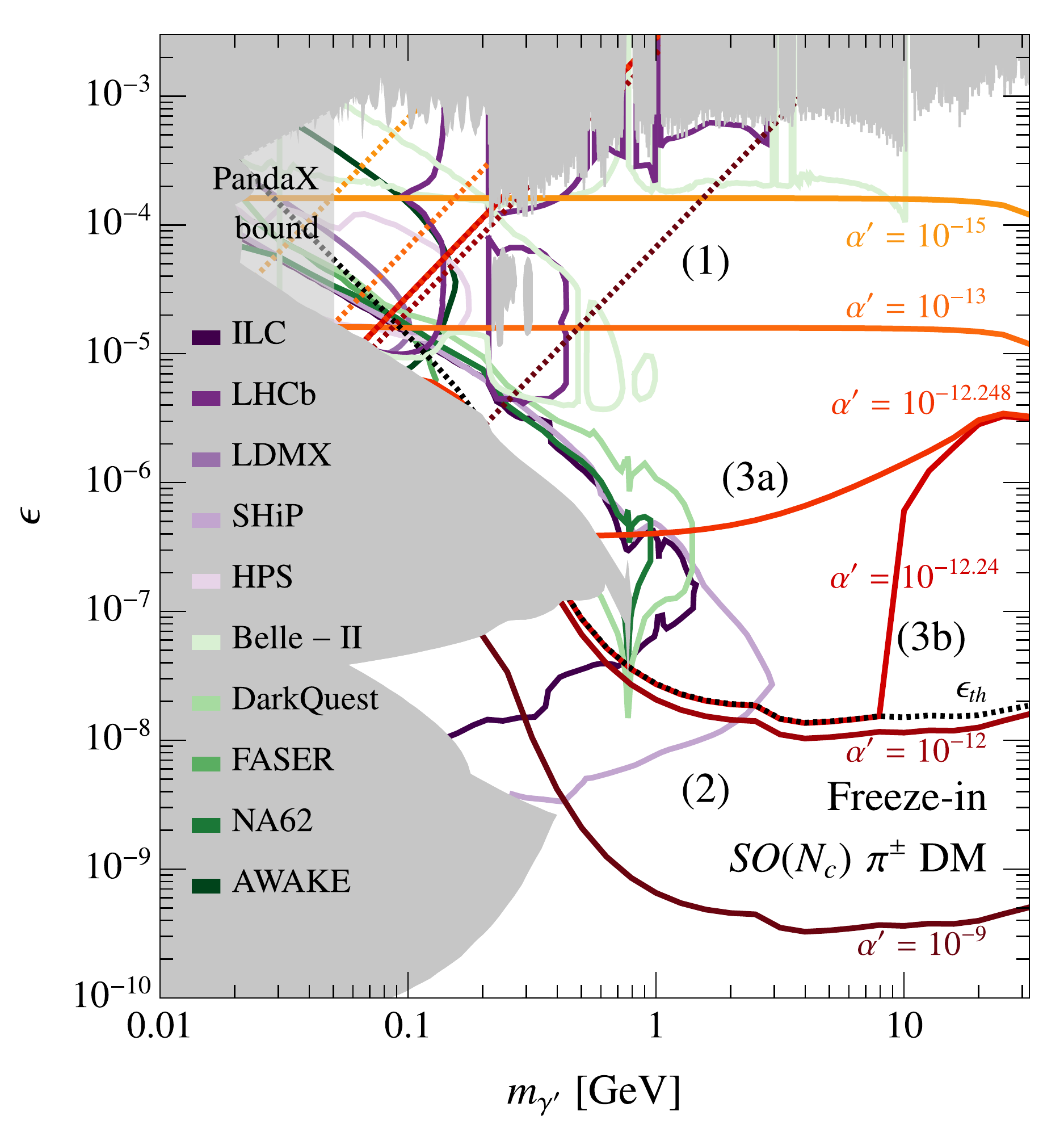}
\caption{\label{fig:SONlightgamFI} Contours of $\alpha '$ on the visibly-decaying dark photon plane $(m_{\gamma '},\epsilon)$ which predict the observed relic abundance via freeze-in in the $SO(N_c)$ model with charged pion dark matter. The constraints and projected sensitivities are the same as in Fig.~\ref{fig:Sp2NlightgamFI}, with a slight change in direct-detection bounds due to all of dark matter being charged in this model (instead of $2/5$). Regions labeled (1), (2), (3a) and (3b) are defined in Appendix~\ref{sec:boltz}. The dark matter mass is assumed to be the best fit value of 16.7~GeV.}
\end{center}
\end{figure*}

In fact, the freeze-in calculations in all regimes of the parameter space are the same as for the previous model and we need not repeat any calculations here. The only distinction we need make is in what values of $\alpha'$ are permitted for this model since direct and indirect detection bounds may apply. For the small values of $\alpha'$ we consider in Fig.~\ref{fig:SONlightgamFI}, we find that indirect detection constraints \cite{Bergstrom:2013jra,Leane:2018kjk} are evaded by orders of magnitude. We likewise assume similarly small values of $\alpha'$ when considering the viable parameter space in Fig.~\ref{fig:heavygamFI} since the heavy-$\gamma'$ freeze-in mechanism does not depend on $\alpha'$. On the other hand, we find that light dark photon masses, $m_{\gamma '} \lesssim 50 \text{ MeV}$, are ruled out by the latest direct-detection results from PandaX-4T~\cite{PandaX-4T:2021bab}. 

\section{Conclusion}
\label{sec:conc}
Over the past two decades, observations of galaxies have revealed a number of persistent and curious features of the gravitationally inferred dark matter distributions. Together, these features could be interpreted as pointing towards a scenario where dark matter self interacts, just like the matter of the visible universe. And, just like the early exploration of the visible universe via nucleon scattering that occurred over seventy years ago, the low-energy kinematics of the dark matter in galaxies means their self interaction is captured extremely (and all too) well by two parameters, making elucidation of the microscopic theory highly challenging. This is explained by Bethe's pioneering work on effective range theory, where the two parameters are identified as the scattering length and the effective range of the interaction.

Applying ERT to dark matter scattering and performing a best-fit scan of ERT parameters and dark matter mass to the currently available galaxy data, we find best-fit values where the dark matter scattering length is anomalously large compared with its Compton wavelength.\footnote{History indeed repeats itself: anomalously large scattering lengths were also observed in nucleon scatterings and prompted Bethe’s original ERT explorations.} Such large values can occur near poles in the scattering amplitudes, corresponding to bound or virtual states. There is another curiosity that arises: the small-scale structure data prefer nucleon-sized dark matter cross sections. Could it be that the dark sector mirrors the visible and contains a QCD-like theory? 

Strongly coupled theories similar to QCD can qualitatively be described by linear sigma models. Indeed, the L$\sigma$M as applied to QCD itself is currently undergoing a revival given recent experimental and lattice data. QCD also provides plenty of inspiration for anomalously large scattering lengths via a number of ``accidents,’' a prominent example being the mass of the $\sigma$ resonance just above the $\pi\pi$ scattering threshold. We have explored L$\sigma$Ms as theories for describing self interactions of dark matter. Here, dark matter particles are pions of a dark gauge group - we consider $SO(N_c)$ and $Sp(2N_c)$ theories. We have translated the L$\sigma$M parameters into the ERT language, quantifying the ``accidents'' that need to occur in the microscopic theory to give rise to the necessary ERT parameters to fit the galaxy data. 

While it is true that many different L$\sigma$Ms are indistinguishable by the ERT scattering parameters describing the low-energy self interactions, the dark matter of the universe must satisfy a host of other criteria which are sensitive to the microscopic parameters. We have studied explicit models with the dark pions of $SO(N_c)$ and $Sp(2N_c)$ gauge theories and mapped out various cosmological constraints as well as those coming from terrestrial experiments. We have discovered viable parameter space in both theories that fits the small-scale structure hints, obtains the correct dark matter abundance via freeze-out or -in, and will be probed at near-future collider and beam dump experiments. 

\acknowledgments
We thank Zhengkang (Kevin) Zhang for collaboration at the early stages of the project, in particular his input to the understanding of effective range theory. We also thank Jenny List and J\'er\^ome Vandecasteele for useful discussions. The works of D.\,K.\ and H.\,M.\ were supported by the Beyond AI Institute at the University of Tokyo. R.\,M.\ is supported in part by the DoE grant DE-SC0007859 and performed part of this work at Aspen Center for Physics, which is supported by National Science Foundation grant PHY-1607611. T.\,M.\ is supported by JSPS KAKENHI grants JP19H05810, JP20H01896, and JP20H00153. The work of H.\,M.\ was in addition supported by the Director, Office of Science, Office of High Energy Physics of the U.S. Department of Energy under the Contract No. DE-AC02-05CH11231, by the NSF grant PHY-1915314, by the JSPS Grant-in-Aid for Scientific Research JP20K03942, MEXT Grant-in-Aid for Transformative Research Areas (A) JP20H05850, JP20A203, and Hamamatsu Photonics, K.K. In addition, D.\,K.\ , T.\,M.\ , and H.\,M.\ are supported by the World Premier International Research Center Initiative (WPI) MEXT, Japan.

\appendix

\section{Linear sigma models (L\texorpdfstring{$\sigma$}{s}Ms) \label{sec:LSM}}
L$\sigma$Ms are renormalizable field theory models that demonstrate specific patterns of spontaneous symmetry breaking where the symmetries are realized linearly. Here we review those associated with vector-like gauge theories, where L$\sigma$Ms are generally strongly coupled and non-renormalizable. While we expect that at tree-level, renormalizable theories  qualitatively capture the nature of the dynamics, in some cases we need non-renormalizable terms to correctly represent the symmetry. 

\subsection{\texorpdfstring{$SU(N_c)$}{SU(Nc)}\label{sec:SULsM}}
Here we consider $SU(N_c)$ gauge theories with $N_f$ quarks in the fundamental representation of both chiralities $q_L$ and $q_R$ (Dirac fermions). In the absence of quark masses, the theory has an $SU(N_f)_L \times SU(N_f)_R \times U(1)_B$ global symmetry. For low $N_f$, it is believed that the dynamics develops a quark bi-linear condensate $\langle q_L^i \overline{q_R}^j \rangle \propto \delta^{ij}$, dynamically breaking the global symmetry to the diagonal subgroup $SU(N_f)_V \times U(1)_B$. Even though this cannot be shown analytically, the supersymmetric versions of the theories with anomaly-mediated supersymmetry breaking suggest this symmetry breaking pattern persists at least up to $N_f = 3 N_c-1$, see~\cite{Murayama:2021xfj,Murayama:2021rak}. 

To realize the symmetries linearly, we need a field $\Sigma$ as an $N_f \times N_f$ matrix that transforms as the bi-fundamental representation
\begin{align}
    \Sigma \rightarrow U_L \Sigma U_R^\dagger \,,
\end{align}
for $U_L \in SU(N_f)_L$ and $U_R \in SU(N_f)_R$. $\Sigma$ is neutral under $U(1)_B$. The general renormalizable Lagrangian is
\begin{align}
    {\cal L}_0 &= {\rm Tr} \partial_\mu \Sigma^\dagger \partial^\mu \Sigma
    + \mu^2 {\rm Tr} \Sigma^\dagger \Sigma
    - \lambda_1 {\rm Tr} \Sigma^\dagger \Sigma \Sigma^\dagger \Sigma
    - \lambda_2 \left( {\rm Tr} \Sigma^\dagger \Sigma \right)^2 .
\end{align}
With $\mu^2 > 0$, the potential has the minimum $\Sigma \propto \mathbf{1}$ with the spontaneous symmetry breaking $SU(N_f)_L \times SU(N_f)_R \rightarrow SU(N_f)_V$. 

However, this Lagrangian has an accidental $U(N_f)_L \times U(N_f)_R$ global symmetry. To avoid the anomalous $U(1)_A$ symmetry, we need to add
\begin{align}
    \Delta {\cal L} &= - \mu_0^{4-N_f} \left( {\rm det} \Sigma + c.c. \right) .
\end{align}
This term is expected to be generated by instantons that break $U(1)_A$ due to the anomaly ({\it i.e.}\/, $\eta'$).

In general, $\Sigma = \frac{1}{2} ({\bm \sigma} + i {\bm \pi})$ is a complex field with $2N_f^2$ degrees of freedom. Among them, $N_f^2 -1$ are massless pseudo-scalar Nambu--Goldstone bosons. On the other hand, there are $N_f^2$  massive scalars, and one massive pseudoscalar. It is interesting to note that this is exactly the bosonic content when supersymmetric $SU(N_c)$ is perturbed by anomaly mediation \cite{Murayama:2021xfj}.

A special case is when $N_f=2$, where the global $SU(2)_L \times SU(2)_R \simeq SO(4)$ symmetry is a real group. Then we can impose the reality condition on $\Sigma$
\begin{align}
    \Sigma^* &= (i\tau_2) \Sigma (-i\tau_2),
    \qquad 
    \tau_2 = \left( \begin{array}{cc}
        0 & -i \\ i & 0
        \end{array} \right) .
\end{align}
As a result, we can write
\begin{align}
    \Sigma = \sigma + i \pi_i \tau_i.
\end{align}
There are only $N_f^2 = 4$ degrees of freedom (namely half as many as other cases), with three massless pseudo-scalar Nambu--Goldstone bosons $\pi_i$ $(i=1,2,3)$ and one massive scalar boson $\sigma$. This is indeed the situation of $SO(4)/SO(3)$ L$\sigma$M with $\phi_i$ $(i=1,2,3,4)$ with
\begin{align}
    {\cal L} &= \frac{1}{2} \partial_\mu \phi_i \partial^\mu \phi_i
    - \frac{\lambda}{4} (\phi_i \phi_i - v^2)^2.
\end{align}
This is the minimal model because the case for $N_f=1$ has no non-anomalous chiral symmetry.

\subsection{Return of the L\texorpdfstring{$\sigma$}{s}M for hadrons\label{sec:SU3LsM}}
Following Nambu and Jona-Lasinio's \cite{Nambu:1961tp,Nambu:1961fr} idea of spontaneous chiral symmetry breaking to explain the properties of pions, Gell-Mann and L\'evy wrote down the L$\sigma$M \cite{Gell-Mann:1960mvl} with $N_f=2$ discussed in the previous subsection. However, the existence of the $\sigma$ state remained controversial throughout the 20th century and the community mostly decided to take the L$\sigma$M as a toy renormalizable model of the correct symmetry-breaking pattern. Consequently, the dynamics of pions have been described by the non-linear sigma model without the $\sigma$ field \cite{Coleman:1969sm} ({\it i.e.}\/, the chiral Lagrangian). But recently, experimental evidence for the $\sigma=f_0(500)$ is established \cite{Pelaez:2015qba} together with the rest of the nonet, and hence the L$\sigma$M has returned as a qualitatively correct description of low-energy QCD.

\begin{figure}[t]
    \includegraphics[width=\textwidth]{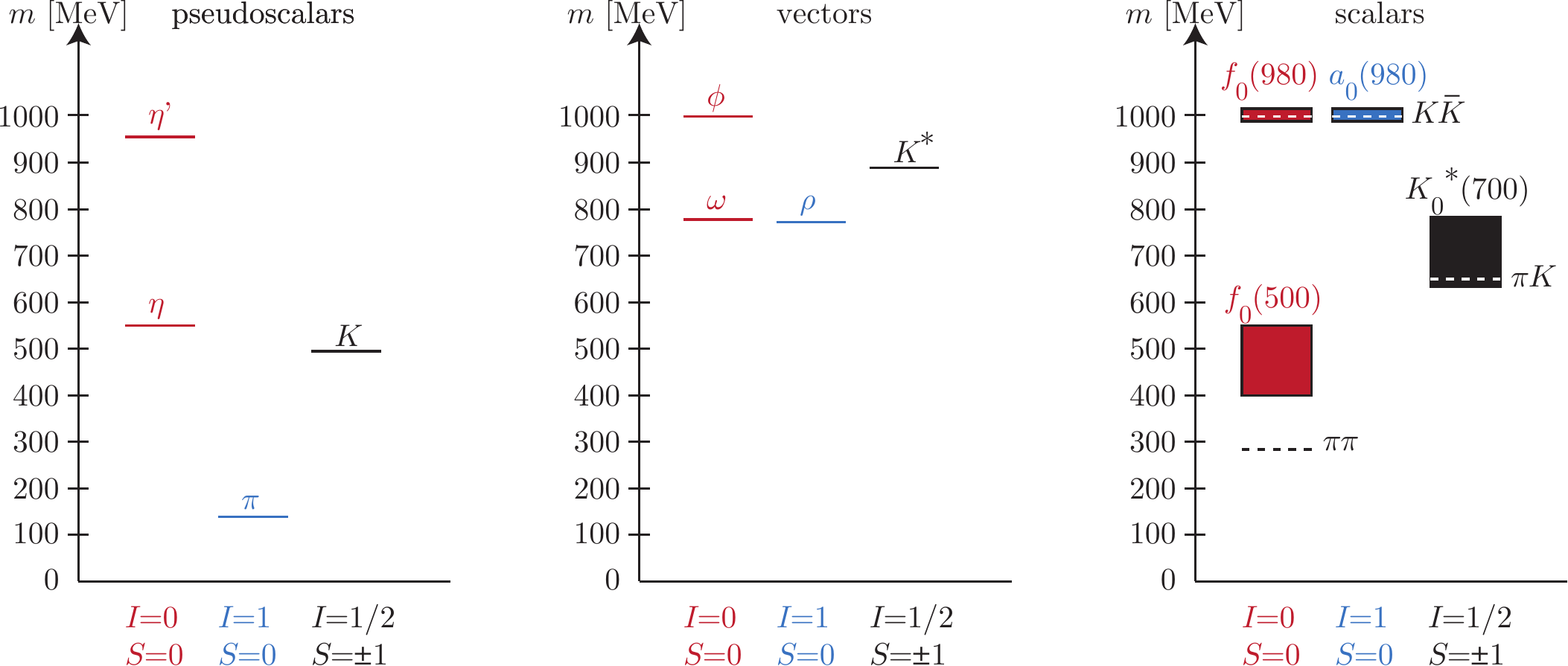}
    \caption{Mass spectra of light pseudoscalar $0^-$, vector $1^-$, and scalar $0^+$ hadrons. The bands indicate the mass range determined in \cite{ParticleDataGroup:2020ssz}. Here, $I$ refers to isospin, and $S$ to strangeness. Pseudoscalar and vector states are qualitatively similar in that the lightest states ($\pi$, $\rho$) form iso-triplets and the heaviest ($\eta'$, $\phi$) iso-singlets with zero strangeness, factoring that pseudoscalar spectrum is more spread out because pions are particularly light as pseudo-Goldstone bosons and the $\eta'$ is heavy due to the chiral anomaly. Scalar states have a qualitatively different mass spectrum, with the iso-triplets $a_0$ at the top and the iso-singlet $f_0(500)$ at the bottom, yet form an $SU(3)$ nonet as evidenced by the Gell-Mann--Okubo relation Eqs.~(\ref{eq:GMO1},\ref{eq:GMO2}). In particular, $f_0(980)$ and $a_0(980)$ are right at the $K\bar{K}$ threshold, pointing to the interpretation as kaon molecules.\label{fig:sigmas} }
\end{figure}

For QCD with the gauge group $SU(3)$ and $N_f=3$, we should have eight Nambu--Goldstone bosons and one massive pseudo-scalar $\eta'$, 
\begin{align}
    {\bm\pi} = 
    \left( \begin{array}{ccc}
        \frac{1}{2} \pi^0 + \frac{1}{2\sqrt{3}} \eta + \frac{1}{\sqrt{6}} \eta'
        & \frac{1}{\sqrt{2}}\pi^+ & \frac{1}{\sqrt{2}} K^+ \\
        \frac{1}{\sqrt{2}} \pi^- & -\frac{1}{2} \pi^0 + \frac{1}{2\sqrt{3}} \eta + \frac{1}{\sqrt{6}} \eta'
        & \frac{1}{\sqrt{2}}K^0 \\
        \frac{1}{\sqrt{2}}K^- & \frac{1}{\sqrt{2}}\overline{K}^0 & - \frac{1}{\sqrt{3}} \eta + \frac{1}{\sqrt{6}} \eta'
    \end{array} \right) \,,
\end{align}
together with a nonet of scalars
\begin{align}
    {\bm\sigma} = 
    \left( \begin{array}{ccc}
        \frac{1}{2} a^0 + \frac{1}{2\sqrt{3}} \sigma + \frac{1}{\sqrt{6}} f_0
        & \frac{1}{\sqrt{2}}a^+ & \frac{1}{\sqrt{2}} \kappa^+ \\
        \frac{1}{\sqrt{2}} a^- & -\frac{1}{2} a^0 + \frac{1}{2\sqrt{3}} \sigma + \frac{1}{\sqrt{6}} f_0
        & \frac{1}{\sqrt{2}}\kappa^0 \\
        \frac{1}{\sqrt{2}}\kappa^- & \frac{1}{\sqrt{2}}\overline{\kappa}^0 & - \frac{1}{\sqrt{3}} \sigma + \frac{1}{\sqrt{6}} f_0
    \end{array} \right).
\end{align}
All of these states are now experimentally well-established with the spectrum in Fig.~\ref{fig:sigmas}, as listed in the Review of Particle Physics \cite{ParticleDataGroup:2020ssz} and the review ``Spectroscopy of Light Meson Resonances'',
\begin{align}
    \sigma &= f_0(500), \\
    (a^\pm, a^0) &= a_0(980), \\
    f_0 &= f_0(980), \\
    (\kappa^\pm, \kappa^0, \bar{\kappa}^0) &= K_0^*(700).
\end{align}
They satisfy the Gell-Mann--Okubo relation (in GeV$^2$):
\begin{align}
     4 m_\kappa^2 
     &= 4\times 0.700^2  = 1.960, \label{eq:GMO1} \\
     m_{a_0}^2 +3m_\sigma^2 &= 0.980^2 + 3\times 0.500^2 = 1.710, \label{eq:GMO2}
\end{align}
which agree within 13\%, suggesting that they indeed form an $SU(3)$ octet. 

\begin{figure}[t]
    \includegraphics[width=\textwidth]{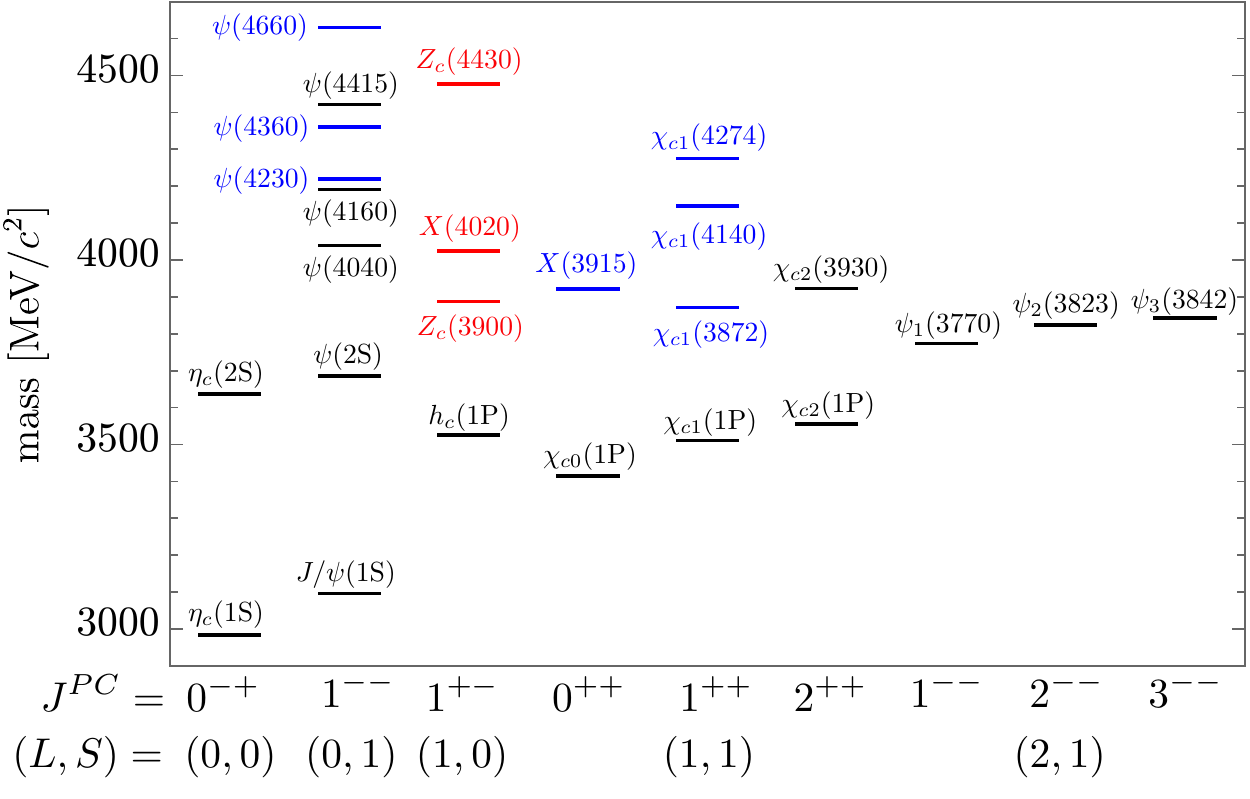}
    \caption{Mass spectra of experimentally well-established states in the Summary Table of ``$c\bar{c}$ mesons (including possibly non-$q\bar{q}$ states)'' from Particle Data Group \cite{ParticleDataGroup:2020ssz}. States shown in black are $c\bar{c}$ charmonium states with configurations indicated by $(L,S)$. On the other hand, states shown in colors are candidates of exotics either as tetraquarks or meson molecules. Those in red are clearly not $c\bar{c}$ as they have charged partners in $I=1$ multiplets, likely $(c\bar{c}u\bar{d}, c\bar{c}(u\bar{u}+d\bar{d})/\sqrt{2}, c\bar{c}d\bar{u})$. Those in blue are $I=0$ yet their properties do not match up with expectations of charmonium states likely $c\bar{c}(u\bar{u}+d\bar{d})/\sqrt{2}$. Those labeled $X$ have their $J^{PC}$ quantum numbers not completely established experimentally. }\label{fig:charmonium}
\end{figure}

The scalar states are unlikely to be understood as $q\bar{q}$ mesons. First of all, the isotriplet states of $q\bar{q}$ mesons would be $u\bar{d}$, $\frac{1}{\sqrt{2}}(u\bar{u}-d\bar{d})$, and $d\bar{u}$ with no strange quark content, and hence are expected to be the lightest in the nonet. This is indeed the case for both pseudoscalars ($\pi$) and vectors ($\rho$), while it is not the case for scalars ($a_0$). Second, there are issues with decay modes as well that $a_0$ decays to $K\bar{K}$ despite being just at the threshold with a minuscule phase space, but not seen in $\pi\pi$. It should be dominated by $s\bar{s}$ content. Third, $q\bar{q}$ mesons have quantum numbers $P=(-1)^{L+1}$, $C=(-1)^{L+S}$, where $L$ is the orbital angular momentum between $q$ and $\bar{q}$ and $S=0,1$ the sum of the spins. Therefore, the $J^{PC} = 0^{++}$ states are required be $p$-wave ($L=1$, $S=1$), which are expected to be heavier than $s$-wave ($L=0$) states. Then we expect closely placed $J=0,1,2$ states similar to $\chi_{c0} < \chi_{c1} < \chi_{c2}$ (see Fig.~\ref{fig:charmonium}) split by spin-orbit coupling and heavier than the $s$-wave $J/\psi$. Instead, we observe both that $f_0(500) < \omega(782)$ ({\it i.e.} the $\omega$ with $L=0$, $S=1$, is heavier than the $\sigma$ with $L=1$, $S=1$) and that $f_0(500) < f_2 (1270) < f_1 (1285)$ ({\it i.e.} the $J=2$ state is lighter than the $J=1$ state).

A possible interpretation of scalar states is tetraquark bound states $(qq)(\bar{q}\bar{q})$, where diquarks $qq$ are spin zero and color anti-triplet and bound to their counterparts anti-diquarks $\bar{q}\bar{q}$ \cite{Jaffe:1976ig,Jaffe:1976ih}. Another possibility is meson molecules $\pi\pi$, $\pi K$, $K\bar{K}$ etc \cite{Weinstein:1982gc}. These do not seem so exotic these days given the observation of hybrids in charm and bottom systems (see Fig.~\ref{fig:charmonium} and also the review ``Non-$q\bar{q}$ Mesons'' in \cite{ParticleDataGroup:2020ssz}). Both interpretations require two quarks and two anti-quarks, and cannot be clearly distinguished; the two interpretations are smoothly connected akin to the BCS-BEC crossover. 

Note that $f_0(980)$ has quantum numbers $I^G(J^{PC})=0^+(0^{++})$ and $a_0(980)$ $1^-(0^{++})$, where the $G$-parity is $(-1)^{L+S+I}$ for $q\bar{q}$ states. The assignment as $q\bar{q}$ states can thus in principle be consistent. However, both states live right on the $K\bar{K}$ threshold; if they were indeed $q\bar{q}$ states, there is no reason why $f_0(980)$ as an $SU(3)$ singlet should have the same mass as $a_0(980)$. The molecules of $K (I=\frac{1}{2})$ and $\bar{K} (I=\frac{1}{2})$ in the $s$-wave yield the same quantum numbers and similar masses for $I=0$ and $I=1$, and hence this interpretation makes sense.  In summary, these states are likely a case of $qq\bar{q}\bar{q}$.

Finally, the fact that the two-flavor case reduces to a triplet of pions and $\sigma$, corresponding to the special case $N_f=2$ with half as many degrees of freedom, is additional evidence for the qualitative correctness of the L$\sigma$M.

\subsection{\texorpdfstring{$SO(N_c)$}{SO(Nc)}\label{sec:SOLsM}}
Here we consider $SO(N_c)$ gauge theories with $N_f$ quarks in the vector representation of fixed chirality (Weyl fermions) $q^i$, $i=1, \cdots, N_f$. In the absence of quark masses, the theory has an $SU(N_f)$ global symmetry. For low $N_f$, it is believed that the dynamics causes a quark bi-linear condensate $\langle q^i q^j \rangle \propto \delta^{ij}$, dynamically breaking the global symmetry to the subgroup $SO(N_f)$. Even though this cannot be shown analytically, their supersymmetric versions with anomaly-mediated supersymmetry breaking suggest this symmetry breaking pattern persists at least up to $N_f = 3 (N_c-2)-1$, see \cite{Murayama:2021xfj,Csaki:2021jax,Csaki:2021xuc}. 

To realize the symmetries linearly, we need a field $\Sigma = \Sigma^T$ as an $N_f \times N_f$ matrix that transforms as the symmetric tensor representation
\begin{align}
    \Sigma \rightarrow U \Sigma U^T
\end{align}
for $U \in SU(N_f)$. The general renormalizable Lagrangian is
\begin{align}
    {\cal L}_0 &= {\rm Tr} \partial_\mu \Sigma^\dagger \partial^\mu \Sigma
    + \mu^2 {\rm Tr} \Sigma^\dagger \Sigma
    - \lambda_1 {\rm Tr} \Sigma^\dagger \Sigma \Sigma^\dagger \Sigma
    - \lambda_2 \left( {\rm Tr} \Sigma^\dagger \Sigma \right)^2 .
\end{align}
With $\mu^2 > 0$, the potential has the minimum $\Sigma \propto \mathbf{1}$ with the spontaneous symmetry breaking $SU(N_f) \rightarrow SO(N_f)$. 

However, this Lagrangian has an accidental $U(N_f)$ global symmetry. To avoid the anomalous $U(1)$ symmetry, we need to add
\begin{align}
        - \mu_0^{4-N_f} \left( {\rm det} \Sigma + c.c. \right) .
\end{align}
This term is expected to be generated by instantons that break $U(1)_A$ due to the anomaly.

In general, $\Sigma$ is a complex field in the symmetric tensor representation of $SU(N_f)$ with $N_f(N_f+1)$ degrees of freedom. Among them, $\frac{1}{2}N_f(N_f+1)-1$ are massless pseudo-scalar Nambu--Goldstone bosons. On the other hand, there are $\frac{1}{2}N_f(N_f+1)$ massive pseudoscalars and one massive scalar. It is interesting to note that this is exactly the bosonic content when supersymmetric $SU(N_c)$ is perturbed by anomaly mediation \cite{Murayama:2021xfj}.

A special case is when $N_f=2$, where the symmetric tensor of the global $SU(2)_L \times SU(2)_R \simeq SO(4)$ symmetry is real $(2,2) = 4$, namely, the vector representation of $SO(4)$. Then we can impose the reality condition on $\Sigma$:
\begin{align}
    \Sigma^* &= (i\tau_2) \Sigma (-i\tau_2),
    \qquad 
    \tau_2 = \left( \begin{array}{cc}
        0 & -i \\ i & 0
        \end{array} \right) .
\end{align}
As a result, we can write (remember $\Sigma = \Sigma^T$)
\begin{align}
    \Sigma = \sigma + i (\pi_1 \tau_1 + \pi_2 \tau_3).
\end{align}
There are only three degrees of freedom (half as many as other cases), with two massless pseudo-scalar Nambu--Goldstone bosons $\pi_i$ $(i=1,2)$ and one massive scalar boson $\sigma$. This is indeed the situation of $SO(3)/SO(2)$ L$\sigma$M with $\phi_i$ $(i=1,2,3)$ with
\begin{align}
    {\cal L} &= \frac{1}{2} \partial_\mu \phi_i \partial^\mu \phi_i
    - \frac{\lambda}{4} (\phi_i \phi_i - v^2)^2.
\end{align}
This is the minimal model because the case for $N_f=1$ has no non-anomalous chiral symmetry.

\subsection{\texorpdfstring{$Sp(2N_c)$}{Sp(2Nc)}\label{sec:SpLsM}}
Here we consider $Sp(2N_c)$ gauge theories with $2N_f$ quarks in the fundamental representation of fixed chirality (Weyl fermions) $q^i$, $i=1, \cdots 2N_f$. In the absence of quark masses, the theory has an $SU(2N_f)_L$ global symmetry. For low $N_f$, it is believed that the dynamics develops a quark bi-linear condensate $\langle q^i q^j \rangle \propto J^{ij}$, dynamically breaking the global symmetry to the subgroup $Sp(2N_f)$. Even though this cannot be shown analytically, their supersymmetric versions with anomaly-mediated supersymmetry breaking suggest this symmetry breaking pattern persists at least up to $N_f = 3 (N_c+1)-1$, see \cite{Murayama:2021xfj}. 

To realize the symmetries linearly, we need a field $\Sigma = -\Sigma^T$ as a $2N_f \times 2N_f$ matrix that transforms as the anti-symmetric tensor representation
\begin{align}
    \Sigma \rightarrow U \Sigma U^T
\end{align}
for $U \in SU(2N_f)$ symmetry. The general renormalizable Lagrangian is
\begin{align}
    {\cal L}_0 &= {\rm Tr} \partial_\mu \Sigma^\dagger \partial^\mu \Sigma
    + \mu^2 {\rm Tr} \Sigma^\dagger \Sigma
    - \lambda_1 {\rm Tr} \Sigma^\dagger \Sigma \Sigma^\dagger \Sigma
    - \lambda_2 \left( {\rm Tr} \Sigma^\dagger \Sigma \right)^2 .
\end{align}
With $\mu^2 > 0$, the potential has the minimum $\Sigma \propto J$ with the spontaneous symmetry breaking $SU(2N_f) \rightarrow Sp(2N_f)$, where $J$ is the $Sp(2N_f)$ group invariant. 

However, this Lagrangian has an accidental $U(2N_f)$ global symmetry. To avoid the anomalous $U(1)$ symmetry, we need to add
\begin{align}
        - \mu_0^{4-N_f} \left( {\rm Pf} \Sigma + c.c. \right) .
\end{align}
This term is expected to be generated by instantons that break $U(1)$ due to the anomaly.

In general, $\Sigma$ is a complex field with $2N_f(2N_f-1)$ degrees of freedom. Among them, $N_f(2N_f-1)-1$ are massless pseudo-scalar Nambu--Goldstone bosons. On the other hand, there are  $N_f(2N_f-1)$ massive scalars and one massive pseudo-scalar. It is interesting to note that this is exactly the bosonic content when supersymmetric $SU(N_c)$ is perturbed by anomaly mediation \cite{Murayama:2021xfj}.

A special case is when $N_f=2$, where the anti-symmetric tensor of the global $SU(4) \simeq SO(6)$ symmetry is real, namely, the vector representation of $SO(6)$. Then we can impose the reality condition on $\Sigma$
\begin{align}
    \Sigma^* &= - J \Sigma J^{-1},
    \qquad 
    J = \left( \begin{array}{cccc}
        0 & 0 & +1 & 0 \\
        0 & 0 & 0 & +1 \\
        -1 & 0 & 0 & 0 \\
        0 & -1 & 0 & 0
        \end{array} \right) .
\end{align}
As a result, we can write (remember $\Sigma = -\Sigma^T$)
\begin{align}
    \Sigma = \left( \begin{array}{cccc}
        0 & \pi_1 & \frac{1}{2} (i\sigma+\pi_5) & \pi_2 \\
        -\pi_1 & 0 & \pi_3 & \frac{1}{2} (i\sigma+\pi_6) \\
        -\frac{1}{2} (i\sigma+\pi_5) & -\pi_3 & 0 & \pi_4 \\
        -\pi_2 & -\frac{1}{2} (i\sigma+\pi_6) & -\pi_4 & 0
        \end{array} \right) .
\
\end{align}
There are only six degrees of freedom (half as many as other cases), with five massless pseudo-scalar Nambu--Goldstone bosons $\pi_i$ $(i=1,\cdots, 5)$ and one massive scalar boson $\sigma$. This is indeed the situation of $SO(6)/SO(5)$ L$\sigma$M with $\phi_i$ $(i=1,\cdots, 6)$ with
\begin{align}
    {\cal L} &= \frac{1}{2} \partial_\mu \phi_i \partial^\mu \phi_i
    - \frac{\lambda}{4} (\phi_i \phi_i - v^2)^2,
\end{align}
because the unbroken group is $Sp(4) \simeq SO(5)$. This is the minimal model because the case for $N_f=1$ has no symmetry breaking as $SU(2)/Sp(2) = \{e\}$ is trivial.

\section{Freeze-in when \texorpdfstring{$m_{\gamma '} < 2m_\pi$}{dark photon lighter than twice dark matter}}
\label{sec:boltz}
\subsection{Boltzmann equations for freeze-in}
In this appendix, we detail the Boltzmann equations relevant for freeze-in when $m_{\gamma '} < 2m_\pi$. For parameters shown in Figs.~\ref{fig:Sp2NlightgamFI} and \ref{fig:SONlightgamFI}, the dark matter does not reach equilibrium with the SM bath. Therefore, the Boltzmann equations are
\begin{align}
\dot{n}_{\text{DM}}+3H n_{\text{DM}}&=
2\left( \avg{\sigma_{SM\rightarrow \text{DM}}v} n^2_{\text{SM}}
+ \avg{\sigma_{\gamma'\rightarrow \text{DM}}v} n^2_{\gamma'} \right), \label{eq:nDM}\\
\dot{n}_{\gamma'}+3H n_{\gamma'}&=
\avg{\sigma_{\gamma'\rightarrow \text{SM}}v}n^{\text{eq}}_{\text{SM}} n^{\text{eq}}_{\gamma'}
+ \langle \Gamma_{\gamma' \rightarrow \text{SM}} \rangle n^{\text{eq}}_{\gamma'}-\avg{\sigma_{\gamma'\rightarrow \text{DM}}v} n^2_{\gamma'}, \label{eq:nDP}
\end{align}
and the inverse reactions can be ignored \cite{Hambye:2019dwd}.  The former equation contains the freeze-in production processes for dark matter, $f\bar{f} \rightarrow \pi^+ \pi^-$ and $\gamma' \gamma' \rightarrow \pi^+ \pi^-$. The latter contains the production of dark photons from $f \gamma \rightarrow f \gamma'$, $\bar{f} \gamma \rightarrow \bar{f} \gamma'$, and $f\bar{f} \rightarrow \gamma \gamma'$ in $\avg{\sigma_{\gamma'\rightarrow SM}v}$ and $\bar{f} f \to \gamma'$ in $\langle \Gamma_{\gamma' \rightarrow SM}\rangle$. The last term contains the depletion of dark photons by their annihilation into dark matter; for the parameters we consider below, this is negligible. Here and below, $f$ refers to SM fermions, and $s_{\text{rad}}$ the entropy density. 

The thermally averaged cross sections in the Boltzmann equations can be written as \cite{Hochberg:2018rjs} 
\begin{align}
\avg{\sigma_{1\,2\rightarrow 3\,4}v}\,n_1^{\text{eq}}n_2^{\text{eq}} &= \frac{1}{g_1 g_2 S}\int d \tilde{p}_{1} d \tilde{p}_{2} d \tilde{p}_{3} d \tilde{p}_{4}
|\mathcal{M}|^2\mathrm{e}^{-\frac{E_1}{T}} \mathrm{e}^{-\frac{E_2}{T}} (2\pi)^4\delta(p_1+p_2-p_3-p_4) \nonumber\\
&=\frac{1}{g_1 g_2 S}\int \frac{ds}{64\pi^4  } T \bar{\beta}_i K_2\left(\frac{\sqrt{s}}{T}\right)s^{\frac{3}{2}} \sigma(s)v  ,
\end{align}
where $g_i$ are the number of spin degrees of freedom in the initial states, $S$ is a symmetry factor to account for cases of identical particles, $s = (p_1 + p_2)^2$, $v$ is the relative velocity, and the Lorentz-invariant phase space integral is given by
\begin{align}
    d \tilde{p} \equiv \frac{d^3\bm{p}}{(2\pi)^3 2E} \ ,
\end{align}
and
\begin{align}
    \bar{\beta}_i = \sqrt{1 - \frac{2(m_1^2+m_2^2)}{s} + \frac{(m_1^2-m_2^2)^2}{s^2}} \ .
\end{align}

Switching to the yield $Y=n/s_{\text{rad}}$, we have $\dot{Y}=\frac{\dot{n}+3Hn}{s_{\text{rad}}}$, and further changing the variable from time to $z=m_{\text{DM}}/T$, we obtain 
$$
\frac{d}{dz} Y=\left(g_* \frac{\pi^2}{90} \right)^{-\frac{1}{2}}  \frac{M_{\text{Pl}}}{m_{\text{DM}}^2} z\,\dot{Y} \,,
$$
where $M_{\text{Pl}}$ is the reduced Planck mass. We list the ingredients for the various $s\sigma v=2s\sigma \beta_i$ in appendix C.

Since the r.h.s. of the equation that sets $Y_{\gamma'}$ depends only on SM particles (neglecting the final term in Eq.~\eqref{eq:nDP} as discussed above), we simply numerically integrate it to obtain $Y_{\gamma'}(z)$. Then we use Eq.~\eqref{eq:nDM} for $Y_{\text{DM}}$ and integrate the r.h.s. using the obtained $Y_{\gamma'}(z)$ to obtain $Y_{\text{DM}}(\infty)$. 

\subsection{Thermal effect\label{sec:thermal}}
Thermal effects can significantly affect dark photon production throughout freeze-in. Following \cite{Hambye:2019dwd}, they can be accounted for via a replacement of the kinetic mixing:
\begin{align}
\epsilon^2 \rightarrow \epsilon_{\text{eff}}^2 
&= \frac{\epsilon^2 m_{\gamma '}^4}{(m_{\gamma '}^2-\text{Re}\Pi_{\gamma})^2+\omega^2(\mathrm{e}^{\frac{\omega}{T}}-1)^2\Gamma^2_{\gamma} } \,.
\end{align}
In the above, $\Pi_{\gamma}$ denotes the in-medium photon self-energy. Regarding this as a thermal mass for the photon, we set it equal to the plasma frequency, $\omega_p$, given by $\omega_{p}^2=\frac{4}{9}\pi \alpha T^2 \sum_f Q_f^2 \simeq (0.26 T)^2$, where the sum runs over light fermion species lighter than $m_{\gamma'}/2$. The second term in the denominator is known as the emission rate, and can be evaluated as in \cite{Redondo:2008ec},
\begin{align}
\omega^2(\mathrm{e}^{\frac{\omega}{T}}-1)^2\Gamma_{\gamma}^2
&=\sum_f\frac{\alpha^4 T^4}{\pi^2}\left(\log \frac{4T\omega}{m_f^2} \right)^2 \,,
\end{align}
where the sum runs over all fermions below $m_{\text{DM}}$. For the production of dark photons through the freeze-in process $f\bar{f} \rightarrow \gamma'$, the dark photon energy is $\omega \sim m_{\gamma'}$, and we use this value in our evaluations. 

\begin{figure*}[t!]
\begin{center}
\includegraphics[width=0.75 \columnwidth]{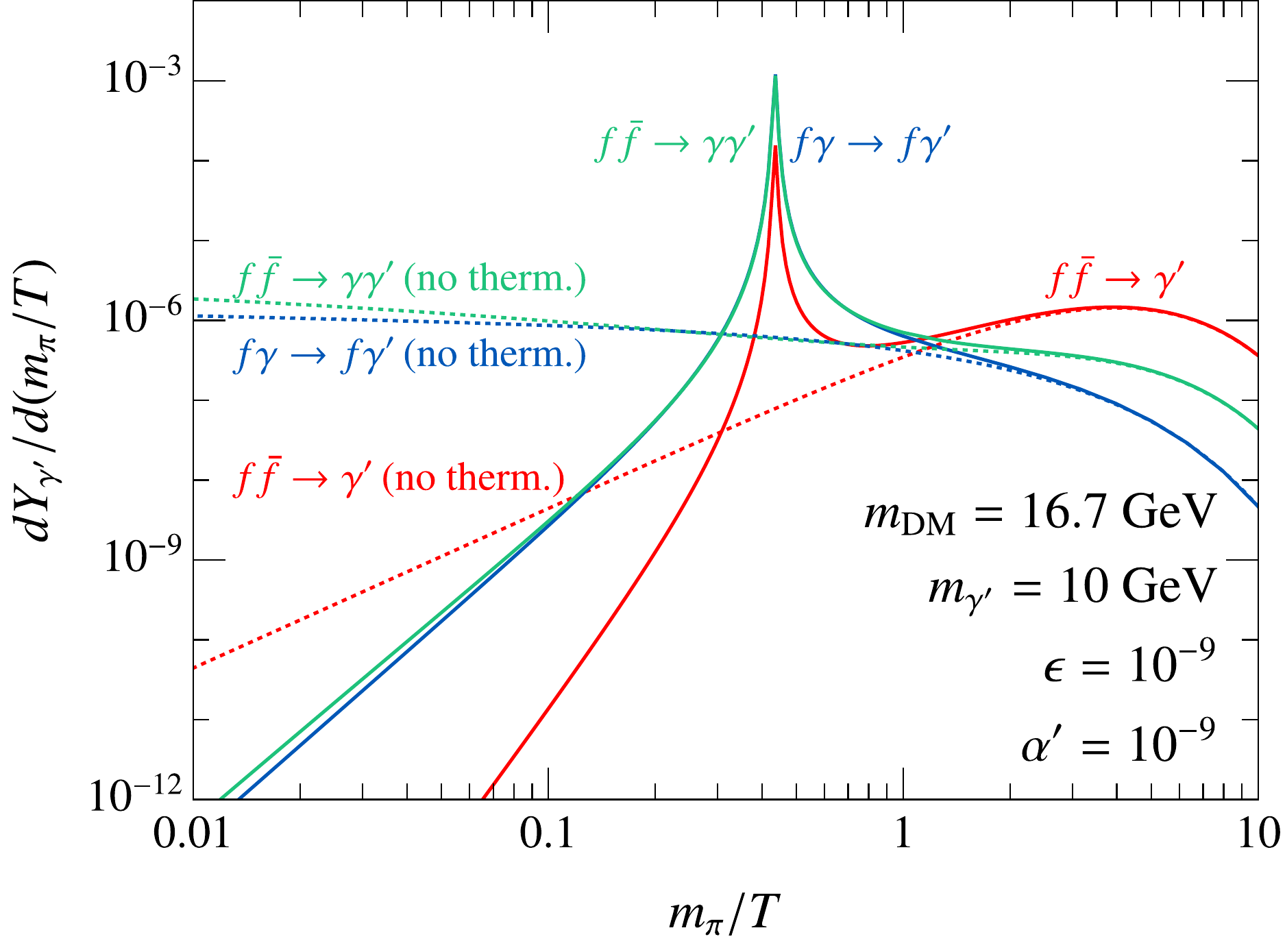}
\caption{\label{fig:dYdz} Contributions to the differential yield $dY_{\gamma '}/d\prn{m_\pi/T}$ as a function of $m_\pi/T$ for the benchmark $m_{\gamma '} = 10 \text{ GeV}$ and $\alpha'=\epsilon=10^{-9}$.  The corresponding dashed lines show the contributions when no thermal effects are included.} 
\end{center}
\end{figure*}

\begin{figure*}[t!]
\begin{center}
\includegraphics[width=0.75 \columnwidth]{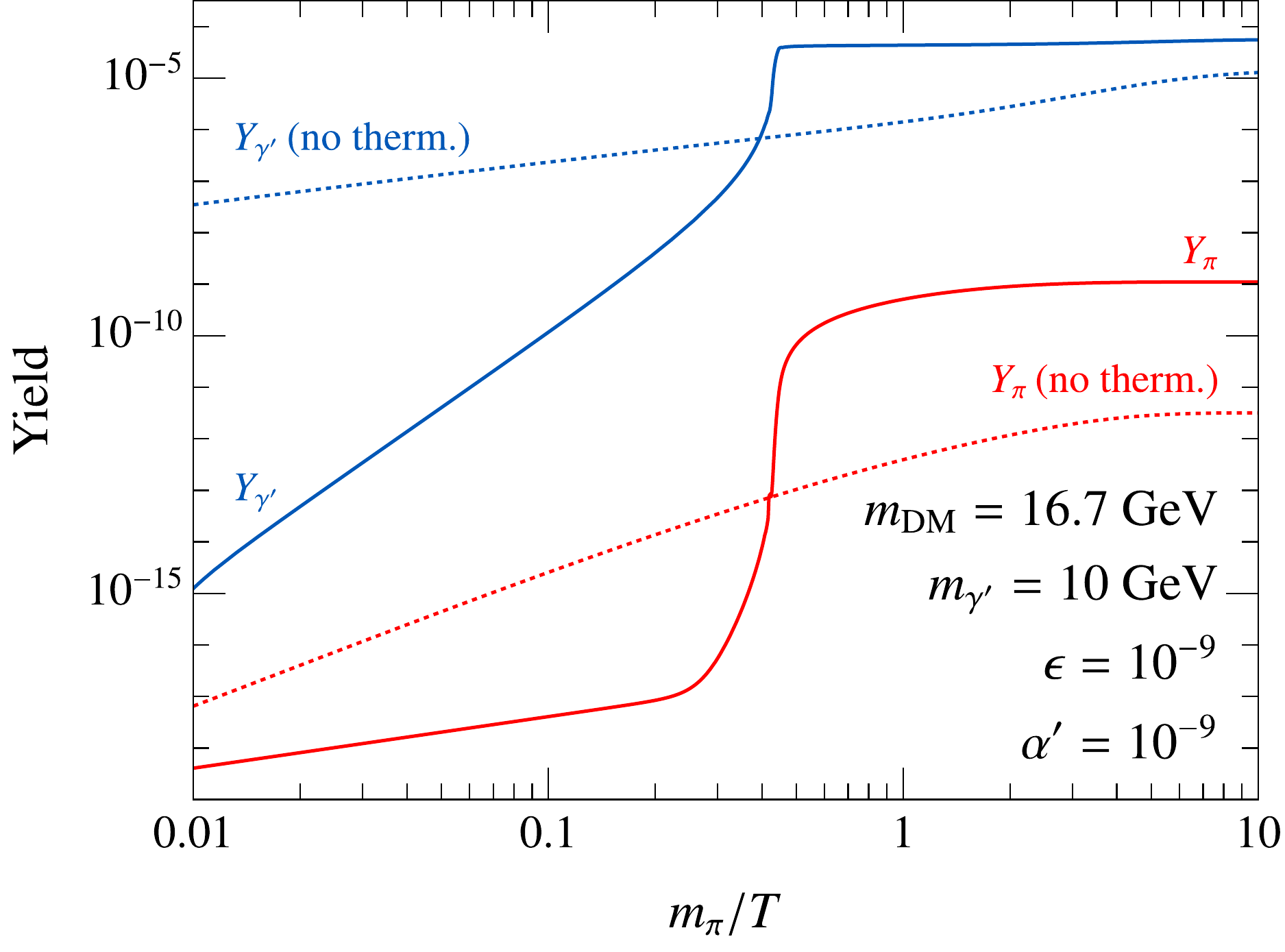}
\caption{\label{fig:yields} The yields of dark photons and pions as a function of $m_\pi/T$ for the benchmark $m_{\gamma '} = 10 \text{ GeV}$ and $\alpha'=\epsilon=10^{-9}$. Also shown are the same yields if no thermal effects are included (dashed).
} 
\end{center}
\end{figure*}

For the best fit $m_{\text{DM}} \sim 16.7$~GeV, we have $\omega_p \sim 1$~GeV at $T \sim \frac{1}{3} m_{\text{DM}}$. Therefore, for $m_{\gamma'} \gtrsim 1$~GeV, $\epsilon_{\text{eff}}$ undergoes a resonance at the temperature $m_{\gamma'} = \omega_p(T_{\text{res}})$. This resonant behaviour is visible in the differential dark photon yields, as shown in Fig.~\ref{fig:dYdz}. The dark photon yield $Y_{\gamma'}$ reaches its asymptotic value quickly after $T_{\text{res}}$, as clearly seen in Fig.~\ref{fig:yields}. $Y_{\gamma'}$ then receives an additional contribution from the inverse decay $f\bar{f} \rightarrow \gamma'$ when $T < m_{\gamma'}$; Fig.~\ref{fig:yields} indeed shows a slight rise in $Y_{\gamma'}$. After the resonance, the enhanced $Y_{\gamma'}$ enables the freeze-in of dark matter through $\gamma'\gamma'\rightarrow \pi^+ \pi^-$, which is also seen in Fig.~\ref{fig:yields}. This corresponds to Region II in~\cite{Hambye:2019dwd} where the sequential freeze-in of dark photons followed by $\gamma'\gamma'\rightarrow \pi^+ \pi^-$ dominates the dark matter production.

On the other hand for $m_{\gamma'} \lesssim 0.3$~GeV, the resonance does not occur and the dark matter production is dominated by the process $f\bar{f} \to \pi^+ \pi^-$. This part of parameter space corresponds to Region Ia in \cite{Hambye:2019dwd} and explains the different plateaus in Figs.~\ref{fig:Sp2NlightgamFI} and \ref{fig:SONlightgamFI}. For $m_{\gamma'} \simeq 0.3\text{--}1$~GeV, we see a transition between two plateaus. 

\subsection{Recipe for the \texorpdfstring{$m_{\gamma'}-\epsilon$}{dark photon mass-epsilon} curves}

Here, we describe how to calculate the $m_{\gamma'}-\epsilon$ curves that give the observed dark matter relic abundance. Let us consider $Y_{\text{DM}}$ when $\alpha'=\epsilon=10^{-9}$ as a benchmark. We calculate the values of $Y_{\text{DM}}^{\text{dir}}$ from direct freeze-in, and $Y_{\text{DM}}^{ \text{seq}}$ from sequential freeze-in. We define $Y_{\text{DM}}^{\text{max}}$ as the yield $Y_{\text{DM}}$ produced from $Y_{\gamma'}^{\text{\text{eq}}}$; this is the maximum yield that can be produced from the sequential freeze-in process. We require two things. First, the observed abundance must come from the two contributions combined, namely the direct and sequential freeze-in yields. Second, the sequential freeze-in contribution must be smaller than $Y_{\text{DM}}^{\text{max}}$ because the dark photon yield is at most $Y_{\gamma'}^{\text{eq}}$. Thus, we can determine $\epsilon$ for different $\alpha'$ by using the various yields:
 \begin{align}
  \left(\frac{\alpha'\epsilon^2}{10^{-27}}\right) Y_{\text{DM}}^{\text{dir}}   +\text{min}\left[
  \left(\frac{\alpha'\epsilon^2}{10^{-27}}\right)^2 Y_{\text{DM}}^{\text{seq}},
  \left(\frac{\alpha'}{10^{-9}}\right)^2Y^{\text{max}}_{\text{DM}}
  \right]=Y_{\text{DM}}^{\text{obs}}=\frac{4.09\times 10^{-10}\,\text{GeV}}{m_{\text{DM}}}\,.
 \end{align}
 We define the quantity $\epsilon_{\text{th}}$ as the value of $\epsilon$ that sets equal the arguments of the min function in the above equation. This is roughly the value of $\epsilon$ at which the dark photon thermalizes before the dark matter freeze-in.

\begin{figure*}[t]
\begin{center}
\includegraphics[width=0.7 \columnwidth]{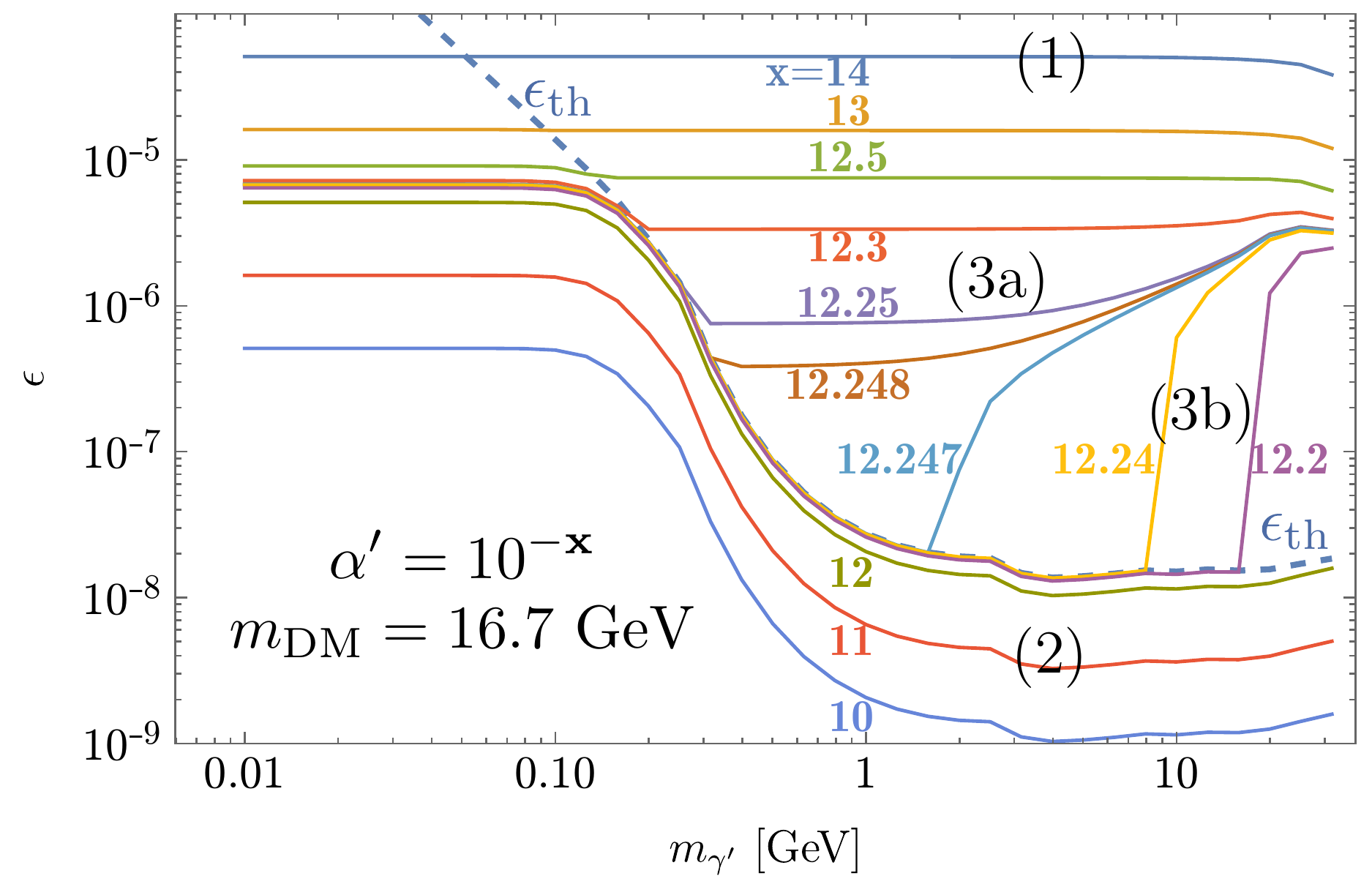}
\end{center}
\caption{\label{fig:transition} Required values of $\epsilon$ for various choices of $\alpha'$ and $m_{\gamma'}$ due to a competition of direct and sequential freeze-in. Above the dashed curve $\epsilon_{\text{th}}$, the dark photon attains a thermal abundance $Y_{\gamma'}^{\text{eq}}$ early enough for the dark matter freeze-in. (1) When $\alpha' < 10^{-12.5}$, dark photon annihilation is too small and direct freeze-in dominates. (2) For $\alpha' > 10^{-12}$, direct freeze-in dominates for $m_{\gamma'} \lesssim 0.1 m_{\text{DM}}$ where $\epsilon_{\text{eff}}$ does not have a resonance, and the required $\epsilon$ is large (left plateau). For larger $m_{\gamma'}$, the resonance leads to a large $Y_{\gamma'}$ and hence dark photon annihilation dominates with smaller $\epsilon$ (middle plateau). (3) For the intermediate $\alpha'$, $Y_{\gamma'}$ gets Boltzmann suppressed for large $m_{\gamma'}$, requiring the direct freeze-in process and hence $\epsilon$ to rise sharply. See text for more details. } 
\end{figure*}

\begin{figure*}[t]
\begin{center}
\includegraphics[width=0.7 \columnwidth]{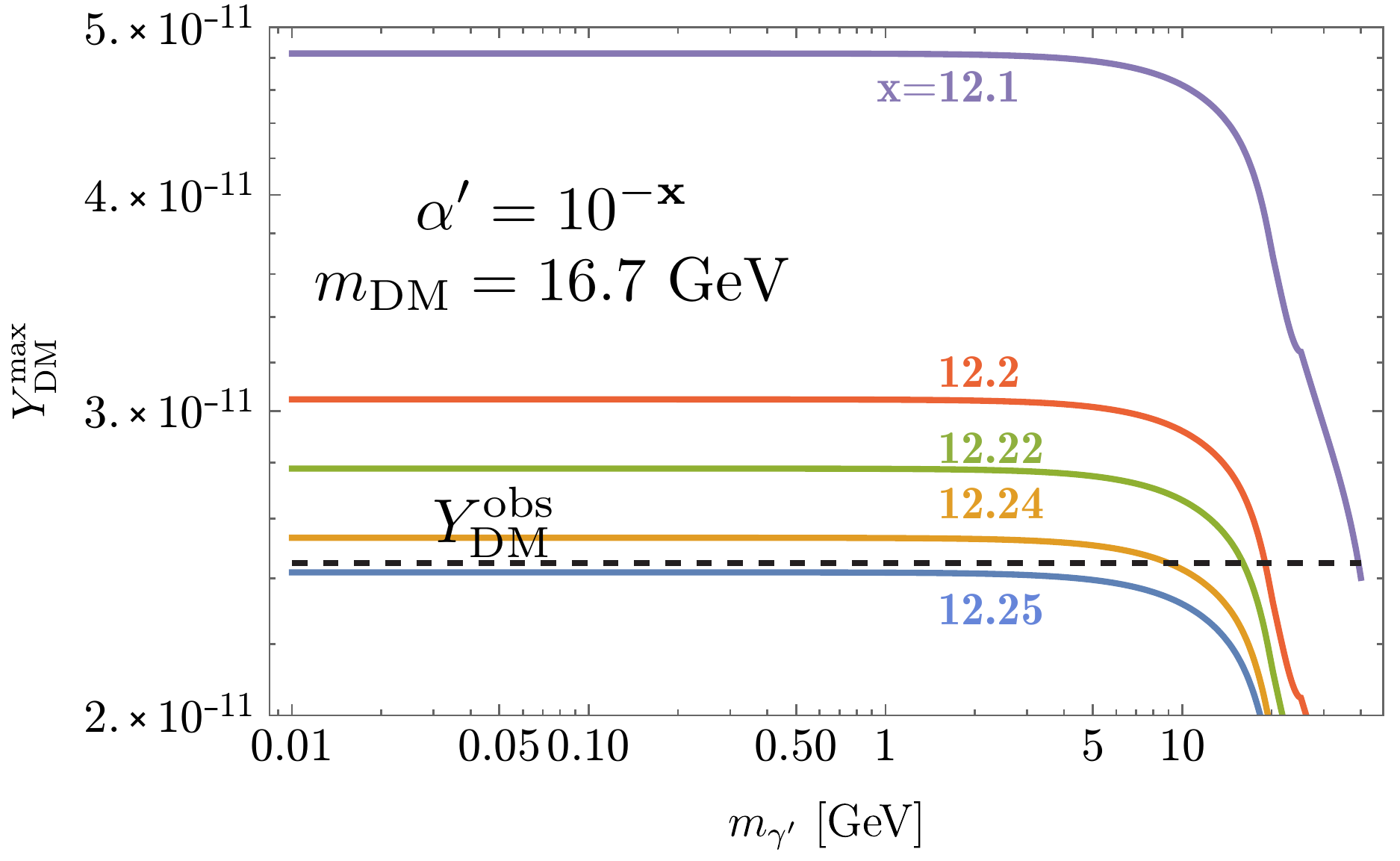}
\end{center}
\caption{\label{fig:threshold} $Y_{\text{DM}}^{\text{max}}$ (freeze-in of dark matter from thermal dark photon $\gamma'\gamma'\rightarrow \pi^+ \pi^-$) for various $\alpha'$ as a function of $m_{\gamma'}$. For $\alpha' \gtrsim 10^{-12.1}$, $Y_{\text{DM}}^{\text{max}}>Y_{\text{DM}}$ for the entire range of $m_{\gamma'}$, and hence requires $\epsilon<\epsilon_{\text{th}}$. For smaller $\alpha'$, however, $Y_{\text{DM}}^{\text{eq}}$ cannot attain $Y_{\text{DM}}$ (black dashed line) when $Y_{\gamma'}$ is Boltzmann suppressed for larger $m_{\gamma'}$, and requires the direct freezein contribution once the curve crosses $Y_{\text{DM}}$. This explains the sharp rise in $\epsilon$ as seen in Fig.~\ref{fig:transition}. For even smaller $\alpha' < 10^{-12.25}$, $Y_{\text{DM}}^{\text{max}}$ is never sufficient and always requires the direct freezein contribution, and $\epsilon$ is much larger than $\epsilon_{\text{th}}$.}
\end{figure*}

We are now able to discuss the behavior of the fixed $\alpha'$ contours in Figs.~\ref{fig:Sp2NlightgamFI} and \ref{fig:SONlightgamFI}, which is different in three regions, denoted (1), (2), (3a), and (3b) in Fig.~\ref{fig:transition}:

\begin{itemize}[leftmargin=*]

\item (1) Dark photon annihilation to create dark matter pairs is too small, suppressed by $\alpha^{\prime 2}$, and the freeze-in process is dominated by the direct freeze-in contribution $\propto \epsilon^2 \alpha'$; hence, $\epsilon \propto (\alpha')^{-1/2}$. As $m_{\gamma'}$ approaches $2 m_{\text{DM}}$, the off-shell propagator is nearly resonant and $\epsilon$ goes down slightly. This corresponds to Region Ia in \cite{Hambye:2019dwd}. 

\item (2) $\alpha' \gtrsim 10^{-12}$. This case is practically made of two plateaus. For light dark photons, $m_{\gamma'} \lesssim 0.10$~GeV, the dark photon resonance does not occur until after dark matter has frozen in; hence, the freeze-in is dominated by direct freeze-in, requiring a relatively large $\epsilon$. On the other hand, for $m_{\gamma'} \gtrsim 1$~GeV, $\epsilon_{\text{eff}}$ fully undergoes the resonance and dark photon production is greatly enhanced. Therefore, freeze-in is dominated by dark photon annihilations. Given the relatively large $\alpha'$, the dark photon yield does not need to reach $Y_{\gamma'}^{\text{eq}}$, requiring a relatively small $\epsilon < \epsilon_{\text{th}}$, which corresponds to region II in \cite{Hambye:2019dwd}. In between, there is a transition. 

\item (3) When $10^{-13} \lesssim \alpha' \lesssim 10^{-12}$, the situation is very interesting. Similar to regions (1) and (2), the region $m_{\gamma'} \lesssim 0.10$~GeV is dominated by direct freeze-in. For larger $m_{\gamma'}$, the resonance starts to kick in and dark photon annihilations become relevant. Then, the region is further subdivided into two cases. (3a) When $10^{-12.248} \lesssim \alpha' \lesssim 10^{-13}$, dark photon annihilations become limited by the thermal dark photon yield, $Y_{\gamma'}^{\text{eq}}$, independent of $\epsilon$ as long as $\epsilon > \epsilon_{\text{th}}$. The rest is made up by the direct freeze-in process, which fixes the required value of $\epsilon$ and there are plateaus. However, for larger $m_{\gamma'}$, the thermal abundance of $Y_{\gamma'}^{\text{eq}}$ gets Boltzmann suppressed. Thus, there is a greater need for the direct freeze-in contribution and the required $\epsilon$ goes up. (3b) When $10^{-12} \lesssim \alpha' \lesssim 10^{-12.247}$, $\alpha'$ is large enough to produce sufficient dark matter even slightly below $\epsilon_{\text{th}}$ for an intermediate range of $m_{\gamma'}$. However, for larger $m_{\gamma'}$, even $Y_{\gamma'}^{\text{eq}}$ is not sufficient to produce dark matter. This is shown in Fig.~\ref{fig:threshold}, which plots $Y^{\text{max}}_{\text{DM}}$ as a function of $m_{\gamma'}$.  When $Y^{\text{max}}_{\text{DM}}$ falls below $Y^{\text{obs}}_{\text{DM}}$, a much larger $\epsilon$ is suddenly required to utilize the direct freeze-in contribution, resulting in a sharp rise in $\epsilon$. 

\end{itemize}

\section{Cross sections}
\label{sec:xsec}
In this appendix, we list some of the cross sections used in the paper for completeness. They always appear in the combination 
\begin{align}
s\sigma v&=
\frac{\bar{\beta}_f}{8\pi} \int \frac{d\Omega}{4\pi}\
\frac{1}{g_1g_2}\sum_{\rm helicities} |\mathcal{M}|^2, \\
\bar{\beta}_f &= \sqrt{1 - \frac{2(m_1^2+m_2^2)}{s} +\frac{(m_1^2-m_2^2)^2}{s^2}}\,
\end{align}
for final state particles of mass $m_1$ and $m_2$. Here, $v$ is the relative velocity.

\subsection{\texorpdfstring{$\text{DM} +\text{DM} \rightarrow f+\bar{f}$}{DM + DM to f + f} (for freeze-out)}

\begin{align}
\sum_{\text{helicities}}|\mathcal{M}|^{2}
&=\frac{(4\pi)^{2}Q^{2}\epsilon_{f}^{2}\alpha\alpha'}{(s-m_{AD}^{2})^{2}+m_{AD}^{2}\Gamma^2}\left[-2(t-u)^2+2s(s-4m_{DM}^2)\right],
\end{align}
\begin{align}
m_{\gamma '}\Gamma (s)
&=\frac{1}{3}\left[\sum_f Q_f^{2}\epsilon_{f}^{2}\alpha (s+2m_f^2)\sqrt{1-\frac{4m_{f}^2}{s}} +\alpha'(s-4m_{\text{DM}}^{2})\sqrt{1-\frac{4m_{\text{DM}}^2}{s}}\right],
\end{align}
where fermion and dark matter should be included only when it is kinematically allowed, {\it i.e.}\/, when the argument of the square root is positive. The width is given as a running width as a function of the dark photon four-momentum squared $s$.

\subsection{Decay \texorpdfstring{$ \gamma ' \rightarrow f+\bar{f}$}{Dark photon to f + f }}

\begin{align}
m_{\gamma '}\Gamma = \frac{1}{3} \beta_f Q_f^2 \epsilon_f^2 \alpha (s+2m_f^2) \,.
\end{align}

\subsection{\texorpdfstring{$\gamma '+\gamma ' \rightarrow \text{DM}+\text{DM} $}{ Dark photon + dark photon to DM + DM}}

\begin{align}
\frac{1}{9}\sum_{\text{helicities}}|\mathcal{M}|^2
=&\frac{1}{9}(4\pi)^{2}\alpha^{\prime 2} \Biggl[\frac{(4m_{\text{DM}}^{2}-m_{\gamma '}^{2})^{2}}{(t-m_{\text{DM}}^{2})^{2}}+\frac{(4m_{\text{DM}}^{2}-m_{\gamma '}^{2})^{2}}{(u-m_{\text{DM}}^{2})^{2}} \nonumber\\
&
 +8+\frac{s}{t-m_{\text{DM}}^{2}}+\frac{s}{u-m_{\text{DM}}^{2}}
+\frac{(s-4m_{\text{DM}}^{2}+m_{\gamma '}^{2})^{2}}{(t-m_{\text{DM}}^{2})(u-m_{\text{DM}}^{2})}\Biggr] \,.
\end{align}

\subsection{\texorpdfstring{$\gamma+f \rightarrow \gamma'+f$}{dark photon+f to photon+f}}
\begin{align}
\frac{1}{4} \sum_{\text{helicities}}|\mathcal{M}|^2
&=\frac{2(4\pi)^{2} Q_{f}^{4}\epsilon^{2}\alpha^{2}}{(s-m_f^2)^2} (2 m_{\gamma '}^2 m_f^2 + m_f^4 - s u + m_f^2 (3 s + u))\nonumber\\
&+ \frac{4(4\pi)^{2}Q_{f}^{4}\epsilon^{2}\alpha^{2}}{(s-m_f^2)(u-m_f^2)}  (-m_{\gamma '}^4 + m_{\gamma '}^2 (-2 m_f^2 + s + u) + m_f^2 (2 m_f^2 + s + u))\nonumber\\
&+\frac{2(4\pi)^{2}Q_{f}^{4}\epsilon^{2}\alpha^{2}}{(u-m_f^2)^2} (2 m_{\gamma '}^2 m_f^2 + m_f^4 - s u + m_f^2 (s + 3 u)) .
\end{align}

\subsection{\texorpdfstring{$f + \bar{f} \to \gamma+\gamma'$}{f+f to photon+dark photon}}
\begin{align}
\frac{1}{4} \sum_{\text{helicities}} |\mathcal{M}|^2
&=\frac{2(4\pi)^{2}Q_{f}^{4}\epsilon^{2}\alpha^{2}}{(t-m_f^2)^2} (-2 m_{\gamma '}^2 m_f^2 - m_f^4 + t u - m_f^2 (3 t + u))\nonumber\\
&+ \frac{4(4\pi)^{2}Q_{f}^{4}\epsilon^{2}\alpha^{2}}{(t-m_f^2)(u-m_f^2)}  (m_{\gamma '}^4 - m_{\gamma '}^2 (-2 m_f^2 + t + u) - m_f^2 (2 m_f^2 + t + u))\nonumber\\
&+\frac{2(4\pi)^{2}Q_{f}^{4}\epsilon^{2}\alpha^{2}}{(u-m_f^2)^2} (-2 m_{\gamma '}^2 m_f^2 - m_f^4 + t u - m_f^2 (t + 3 u)) . 
\end{align}

\bibliographystyle{JHEP}
\bibliography{ref}
\end{document}